\documentclass[a4paper,fleqn,usenatbib]{mnras}

\usepackage{newtxtext,newtxmath}
\hypersetup{draft}

\usepackage[T1]{fontenc}
\usepackage{ae,aecompl}
\usepackage{placeins}
\usepackage{float}

\usepackage[dvipsnames]{xcolor}
\usepackage{float}
\usepackage{graphicx}	% Including figure files
\usepackage{amsmath,siunitx}	% Advanced maths commands
\usepackage{mathtools}
\usepackage{tabularx}
\usepackage{leftidx}
\usepackage{afterpage}
\usepackage{cprotect}
\usepackage{subcaption}
\usepackage{tikz}
\usepackage{mwe,tikz}
\usepackage[percent]{overpic}

\graphicspath{{plots/}}

\newcommand{\kms}{\hbox{km\,s$^{-1}$}}

\newcommand{\Rsun}{\hbox{$R_\odot$}}

\newcommand{\Lsun}{\hbox{$L_\odot$}}
\newcommand{\Lstar}{\hbox{$L_\mathrm{\ast}$}}
\newcommand{\Msunyr}{\hbox{$M_\odot \,\hbox{yr}^{-1}$}}
\newcommand{\Mdot}{\hbox{$\dot M$}}

\newcommand{\cmfgen}{\textsc{cmfgen}}

\newcommand{\sivdoub}{\ion{S}{IV} $\lambda \lambda$1062, 1073}
\newcommand{\phosdoub}{\ion{P}{V} $\lambda\lambda$1117, 1128}
\newcommand{\osixdoub}{\ion{O}{VI} $\lambda\lambda$1031, 1037}
\newcommand{\nvdoub}{\ion{N}{V}  $\lambda\lambda$1238, 1242}

\title[Shell Wind Model of AzV83]{Using Shell Models to Investigate Clumping in the Wind of the O7Iaf+ Supergiant AzV83}

\author[Brian L. Flores and D. John Hillier]{
	Brian L. Flores$^{1}$ and 	D. John Hillier$^{1}$
	\\
	$^{1}$ Department of Physics and Astronomy \& Pittsburgh Particle Physics, Astrophysics and Cosmology Center (PITT PACC), \\ \hspace{1cm}  University of Pittsburgh, 3941 O'Hara Street,  Pittsburgh, PA 15260, USA \thanks{blf40@pitt.edu}
	}

\date{Accepted XXX. Received YYY; in original form ZZZ}

\pubyear{2015}

\begin{document}
\label{firstpage}
\pagerange{\pageref{firstpage}--\pageref{lastpage}}
\maketitle

\begin{abstract}
Hot massive stars exhibit strong stellar winds  that enrich the surrounding interstellar medium and affect the stars' evolution. However, the winds are inhomogeneous (clumped) and are difficult to model in radiative transfer codes. To produce more realistic spectra many codes use a volume-filling factor approach to incorporate the effects of clumping. While this approach is convenient it is simplistic. We introduce an alternative approach to incorporate clumping by assuming the wind is composed of dense spherical shells. Using this approach in the radiative transfer code \textsc{cmfgen} we produce synthetic spectra for AzV83, an O7Iaf+ supergiant located in the Small Magellanic Cloud. The spectrum of AzV83 is rich in both photospheric and wind features, making it an ideal candidate with which to investigate the physical characteristics of stellar winds. Synthetic spectra are compared to the star's observed spectrum to better characterize the influence of clumped winds on spectral features, and to better understand the limitations of the volume-filling factor approach. The approach using spherical shells yields similar wind parameters to those obtained using the volume-filling factor approach although a slightly higher mass-loss rate is required to fit H$\alpha$. As expected, the interclump medium in the model with shells allows the high ionisation resonance transitions of \ion{N}{V} and \ion{O}{VI} to be fitted using $L_{\rm X-ray}/L_{\rm Bol} \approx 10^{-7}$ which is typically observed for O stars, and which is a factor of ten lower than needed with the volume-filling factor approach.
\end{abstract}


%%%%%%%%%%%%%%%%%%%%%%%%%%%%%%%%%%%%%%%%%%%%%%%%%%

%%%%%%%%%%%%%%%%% BODY OF PAPER %%%%%%%%%%%%%%%%%%

\section{Introduction}\label{sec_intro}
Radiatively driven winds of hot, massive stars have been shown to be inherently unstable, leading to the formation of an inhomogeneous structure called \textit{clumps} \cite[see overview in][]{Puls2008}. Further, synthetic spectra from radiative transfer codes that do not correct for clumping have been shown to be inconsistent with observations -- features that depend linearly on density (e.g., electron scattering wings in Wolf-Rayet (WR) stars and many UV P~Cgyni profiles in O stars) cannot be fitted with the same mass-loss rate that is needed to match most  emission features (e.g., H$\alpha$  emission in O stars, most WR emission lines) \citep{Hillier1991,Hamann1998,Hillier2003}. In O stars, the occurrence of P~Cygni profiles of so-called super-ions (\ion{O}{VI} $\lambda\lambda 1031, 1037$ and \ion{N}{V} $\lambda\lambda 1238, 1242$) are also inconsistent with smooth homogeneous winds \citep[e.g.][]{Pauldrach1994,Zsargo2008}. In the Large Magellanic Cloud (LMC), phosphorus abundances derived from spectroscopic studies of O stars may differ from the global abundance trends if clumping is ignored \citep{Massa2003}. Consequently the analysis of massive star spectra requires an allowance for the inhomogeneous properties of their winds, especially if our goal is to derive accurate stellar parameters such as mass-loss rates and abundances. 

The volume-filling factor (VFF) approach is the principal method adopted to treat clumping. The winds are assumed to be composed of small, optically thin clumps that occupy a fraction $f_\mathrm{v}$  of the wind, and which are embedded in a void background. Typically it is assumed the wind is homogeneous close to the star (i.e., below the sonic point), becoming progressively clumpy at larger radii. In this approach the clump density ($\rho_\mathrm{cl}$) is given by $\langle\rho\rangle/f$ where $\langle\rho\rangle$ is the local mean wind density. The opacity and emissivity, computed
using the clump density, are scaled by $f_\mathrm{v}$ because along any given ray, only a fraction $f_\mathrm{v}$ of the ray will pass through a clump. With these assumptions, $\rho$-dependent features remain unchanged while $\rho^2$-dependent features are enhanced by the inverse of $f_\mathrm{v}$. As a consequence the mass-loss rate required to fit a density-squared diagnostic, such as the H$\alpha$ emission, is scaled by a factor of $\sqrt{f_\mathrm{v}}$ (and hence reduced). Typical estimates for $\sqrt{f_\mathrm{v}}$, which are dependent on the type of emission-line star, are 0.01 to 0.5  \citep{Bouret2003,Bouret2005,Crowther2002,Hamann1998,Hillier2003,Najarro2001,Puls2003,Puls2006}.

While the use of the VFF approach in radiative transfer codes has been overwhelmingly successful in synthesizing spectra that are comparable to observations, it is simplistic \citep{1996LIACo..33..509H,Oskinova2007,Owocki2008,Zsargo2008}. First, the clumps are assumed to be optically thin. While this is likely to be valid for the continuum, theoretical calculations suggest that clumps have sizes similar to the Sobolev length, making the assumption invalid for lines. Second, it neglects the interclump medium (ICM) which is pertinent to resolving the discrepancy between the observed and theoretical strength of some UV resonance lines of super-ions \citep{Zsargo2008}, and which can also be important for achieving black saturated absorption in other UV profiles \citep{Sundqvist2010,Surlan2012}. Third, it does not allow for effects of porosity and vorosity (porosity in velocity) \citep{Owocki2008,Oskinova2007}. Finally, it does not account for the presence of non-monotonic velocity fields.
Ideally, clumpy winds should be modeled with a sophisticated 3D hydrodynamical radiative transfer code. This, however,  is extremely difficult and computationally too expensive to fully realize. The 2D structure of winds, and their influence on line profile variability, was  explored in a series of papers by Dessart and Owocki (\citeyear{DO02_3D,DO02_clump,DO05_var,DO05_line_force}). They showed, for example, that the variability was consistent with blobs having a lateral extent of $3^{\circ}$, although the radial velocity dispersion they obtained from their 1D models was too low to explain the observed variability behaviour at the edge of the line profile. In 2D simulations of the line-driven instability \cite{DO03_2Dsims} showed that clumps exhibited very little lateral extent -- shells are broken up by thin-shell Rayleigh-Taylor instabilities. They note, however, that the diffuse radiation field, neglected in their simulations, may set a minimum lateral scale.

Several papers have explored other methods of circumventing this roadblock by artificially constructing clumpy wind models. These include: composition of randomly distributed spherical clumps on a non-void background \citep{Surlan2012}; a stochastically distorted smooth, stationary, and spherically symmetric wind model \citep{Sundqvist2010,Sundqvist2011}; and a 2D statistical distribution of discrete spherical shell fragments on a void background \citep{Oskinova2004}. These studies revealed physical insights into clump-sensitive spectral features, such as the influence  of macroclumping\footnote{Macroclumping refers to clumping on a size scale larger than
the mean-free path of a photon. In such cases the clump shape and structure  will also influence its emergent spectrum.
This contrasts with the VFF approach in which the clump shape is irrelevant. The distinction between
microclumping and macroclumping is artificial since there will be a distribution of clump sizes, and since the photon mean-free path is frequency dependent.} on the shape and saturation of UV resonance lines, the emergent X-ray flux, and optical emission line fluxes. However these approaches are restricted to a small set of non-LTE problems (i.e., neglecting non-LTE effects within optically thick clumps or decoupling the emission term from absorption in the radiative transfer equation in order to acquire its formal solution), restricting the kind of winds that can be modeled and studied. 

To push forward our endeavour to treat clumpy winds more realistically and test the validity of the VFF approach, we treat a clumpy wind as a series of dense, spherically symmetric shells embedded on a non-void interclump medium  in the 1D radiative transfer code \textsc{cmfgen}. The advantages of this method are severalfold: It allows us to quantitatively investigate clump-sensitive features over the full observable spectrum and it allows for an ICM. Importantly, we can solve the non-LTE problem for all spectral features. Further, it provides results -- such as ionisation structures and recombination rates of each species --  that can be compared with those from the VFF approach. 

For our first investigation we chose to model spectra of AzV83, an O7Iaf+ supergiant in the Small Magellanic Cloud (SMC), which has previously been modeled by \cite{Hillier2003}. Its spectra shows evidence of a relatively strong mass loss, similar to a WR star, as well as a wealth of photospheric features. Most other massive stars would either have a weak mass loss resulting in few wind features, or winds so opaque that we can no longer see photospheric features. Thus with AzV83 we can study the photosphere, the photosphere/wind transition zone, and the wind.

This paper is organized as follows. In Section \ref{sec_obs_mod} we briefly present spectroscopic observations of AzV83 and outline the two methods, VFF and Shell, for treating clumpy winds. In Section \ref{sec_results} we compare and discuss the atmospheric and wind models constructed using the two methods, emphasizing their differences in ionisation structure and wind lines. In Section \ref{sec_discuss} we outline the advantages of the Shell approach. and its limitations.  Conclusions and future work are presented in Section \ref{sec_conclusion}.

%%%%%%%%%%%%%%%%%%%%%%%%%%%%%%%%%%%%%%%%%%%%%%%%%%%%%%%%%%%%%%%%%%%%%
%%%%%%%%%%%%%%%%%%%%%%%%%%%%%%%%%%%%%%%%%%%%%%%%%%%%%%%%%%%%%%%%%%%%%
\section{Observations and Modeling}\label{sec_obs_mod}
\subsection{Observational Data}\label{subsec_obs}

Our observational data set of AzV83 consists of high-resolution far-UV \textit{FUSE} spectra, ultraviolet spectra obtained with the \textit{Hubble Space Telescope} Space Telescope Imaging Spectrograph (STIS), and ground-based red and blue echelle spectroscopy. The data set is that used by \cite{Hillier2003} but is  summarized here for completeness.
The observed \textit{FUSE} spectrum of AzV83 was originally provided by the principal investigator J. Hutching (Program, P117). Their observations were done with a $30\arcsec\times30\arcsec$ aperture  for a total exposure time of 4010 s. The spectrum spans a wavelength range of 905 -- \SI{1190}{\angstrom} and has a resolving power \textit{R} = 20,000. The STIS data were obtained with the far-UV Multi-Anode Microchannel Array (MAMA) detector in the E140M model with a $0\farcs2\times0\farcs2$ entrance aperture. The spectrum spans a wavelength range of 1150 -- \SI{1700}{\angstrom} with an effective spectral resolving power of \textit{R} = 46,000. The blue echelle spectrum was obtained with the Cassegrain echelle spectrograph (CASPEC) at the 3.6 m telescope at ESO while the red echelle spectrum was obtained with the University College London echelle spectrograph (UCLES) at the coud\'{e} focus of the 3.9 m Anglo-Australian Telescope (AAT). The blue and red spectra span a wavelength range of 3910 -- \SI{5170}{\angstrom} and 4586 -- \SI{8293}{\angstrom}, respectively, at a resolution of 12\,\kms.  For a more in-depth description of the telescopes' specifications, refer to \cite{Walborn2000}  (ultraviolet and blue/red echelle spectra), and \cite{Moos2000} and \cite{Sahnow2000} (\textit{FUSE} spectra). Some parts of the observed spectra had to be locally renormalized since their continuum was above unity. Lines affected by this include \sivdoub, \ion{He}{II} $\lambda$1640, and H$\beta$ $\lambda$4861.

\begin{figure} 
	\includegraphics[width=0.95\columnwidth]{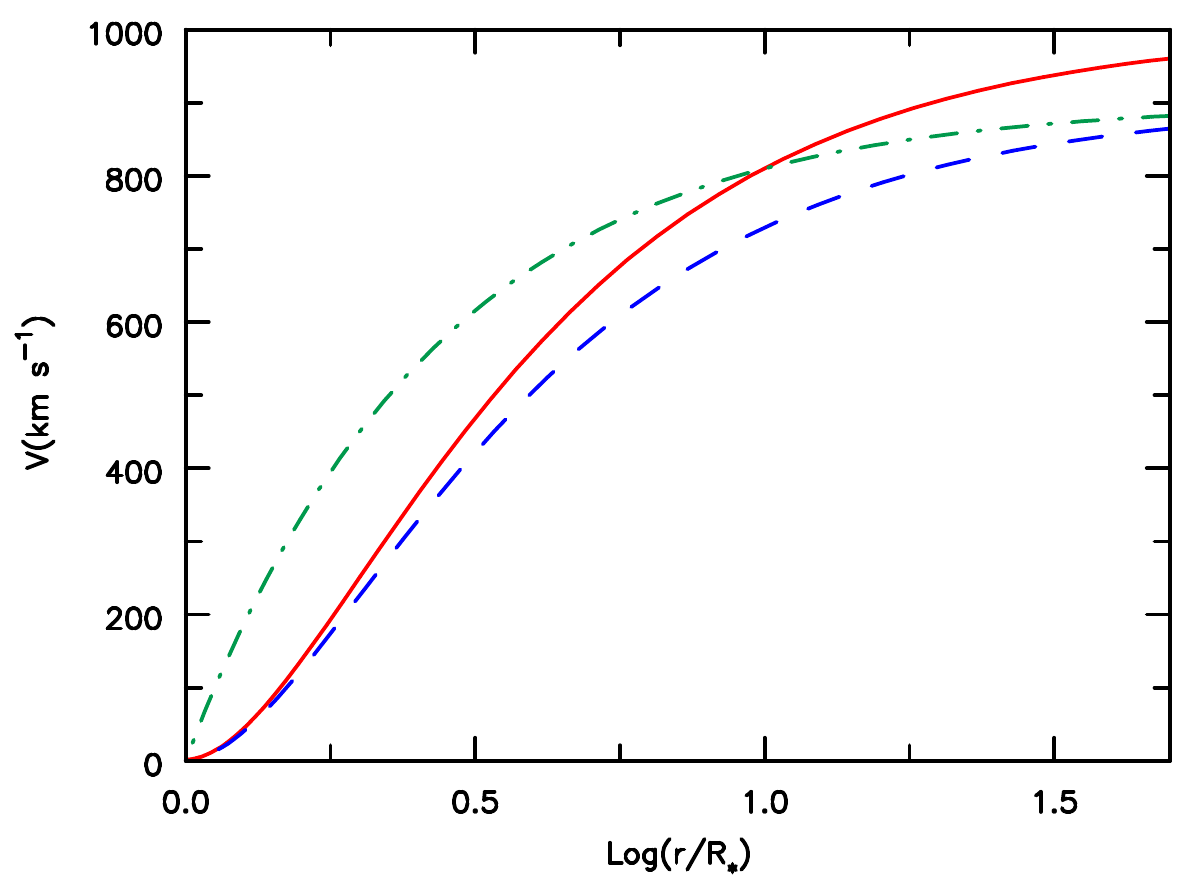} 		\caption{Illustration of $\beta$-velocity law for two different values of terminal velocity $V_\mathrm{\infty}$,  900\,\kms\ (\textit{blue, dash curve}) and 1000\,\kms\ (\textit{red, solid curve}), used in our models. For these plots, $\beta$ = 2.0 and $V_\mathrm{Phot}$ = 10 \kms. A $\beta$-velocity law using $\beta$ = 1.0 and $V_\mathrm{\infty} = 900$\,\kms\ (\textit{green, dash-dot curve}) is shown for comparison.
	}
	\label{fig_vel_law}
\end{figure}

\begin{table}%[h]
	\centering
	\caption{Atmospheric and wind parameters that lead to the best agreement with the observations of AzV83. Atmospheric values listed under the VFF model are also shared by the Shell model. Some values listed below are unique to the Shell model; these parameters are used in the construction of a Gaussian distribution function which is used as our shell profile.}

	\begin{tabularx}{\columnwidth}{l*{2}{X}} 
		\hline
		\textbf{Parameter} & \textbf{VFF model} & \textbf{Shell model}		\\ \hline
		Number of Depth Points (ND)          &\textasciitilde 60 &\textasciitilde 550  \\[2.5pt]
		$L_\mathrm{*}        ~(10^5$ \Lsun)         & 4.65              &  4.65 \\[2.5pt]
		${R_\mathrm{2/3}} ^a ~$(\Rsun)              & 20.68             &  20.69\\[2.5pt]
		$T_\mathrm{eff}      ~$(kK)                 & 33.13             &  33.12\\[2.5pt]
		$L_\mathrm{X}        ~(10^{-7}$ \Lstar)     & 1.0               &  1.0\\[2.5pt]		
		$ \dot{M}(10^{-7} \,$\Msunyr)            & 8.0 & 11.0  \\[2.5pt]
		$V_\mathrm{Phot}$(\kms)                  & 10.0& 10.0  \\[2.5pt]
		$V_\mathrm{\infty}$(\kms)                & 900 & 1000  \\[2.5pt]
		$\beta                               $   & 2.0 & 2.0     \\[2.5pt]
		${f_\mathrm{v}}^b                    $   & 0.1 & \textasciitilde 0.1 \\[2.5pt]
		
		$\text{FWHM}      ~(\leq1.0)  $          & --  & 0.05  \\ [2.5pt] 
		$R_\mathrm{\text{scl}}   ~(>1.0)     $   & --  & 2.00  \\ [2.5pt]
		$V_\mathrm{\text{onset}}$(\kms)          & --  & 27.0  \\ [2.5pt]
		$V_\mathrm{\text{clump}}$(\kms)          & --  & 100   \\ [2.5pt] 
		ND\textsubscript{shell}                  & --  & 35    \\ [2.5pt]
		
		\hline
	\end{tabularx}
	\begin{tabular}{l}
		\textbf{Notes.}\\
		$ \leftidx{^a}R_\mathrm{2/3}$ corresponds to the radius at which the Rosseland optical depth is 2/3.\\
		$ \leftidx{^b}f_\mathrm{v}$ is the volume-filling factor; for the Shell model, this is calculated \\
		$~~~$with $\rho_\mathrm{cl}=\langle\rho\rangle/f_\mathrm{v}$, where $\langle\rho\rangle$ is the smooth wind density. This term \\
		$~~~$is dependent on FWHM. 
	\end{tabular}
	\label{table:Param}
\end{table}

\begin{figure} %shell profile
	\includegraphics[width=0.95\columnwidth]{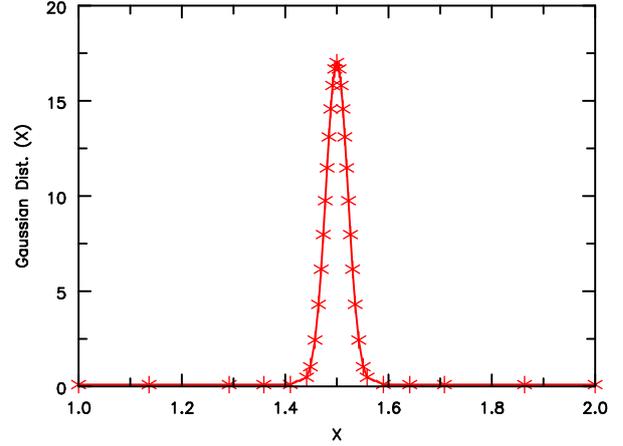}
	\caption{Illustration of a shell profile with the following parameters: $R_\mathrm{scl}=2.0$, FWHM = 0.05, and ND\textsubscript{shell} = 35. The shell profile requires additional grid points (\textit{red, asterisk}) to resolve it.}
	\label{fig_shell_prof}
\end{figure}

%%%%%%%%%%%%%%%%%%%%%%%%%%%%%%%%%%%%%%%%%%%%%%%%%%%%%%%%%%%%%%%%%%%%
\subsection{Atmospheric and Wind Modeling}\label{subsec_mod}

To study the two methods, we performed spectral analysis of atmospheric models calculated using the non-LTE line-blanketed multi-purpose atmospheric code \textsc{cmfgen} \citep{Hillier1998}. \textsc{cmfgen} solves the radiative transfer equation for spherical geometry in a comoving frame in conjunction with the equations of statistical and radiative equilibrium. We do not solve the hydrodynamic equations for the wind -- instead we assume that the wind's velocity field is described by the $\beta$-velocity law. Based on earlier work, we adopt $\beta = 2$; the resulting velocity field is plotted in Fig. \ref{fig_vel_law}. We construct two sets of models: one using our Shell approach and the other using the commonly accepted VFF approach for treating clumping.\footnote{The Shell option will become publicly available in a future update for \textsc{cmfgen} on \url{http://www.pitt.edu/~hillier/}. Register to receive emails about updates to \textsc{cmfgen}.} Since our goal is to understand the limitations of the VFF approach, both sets of models adopt similar atmospheric parameters (listed in Table \ref{table:Param}), atomic models and data (Appendix~\ref{atomic_data}), and surface chemical abundances (Appendix~\ref{abundances}).

\begin{figure}
	\includegraphics[width=0.95\columnwidth]{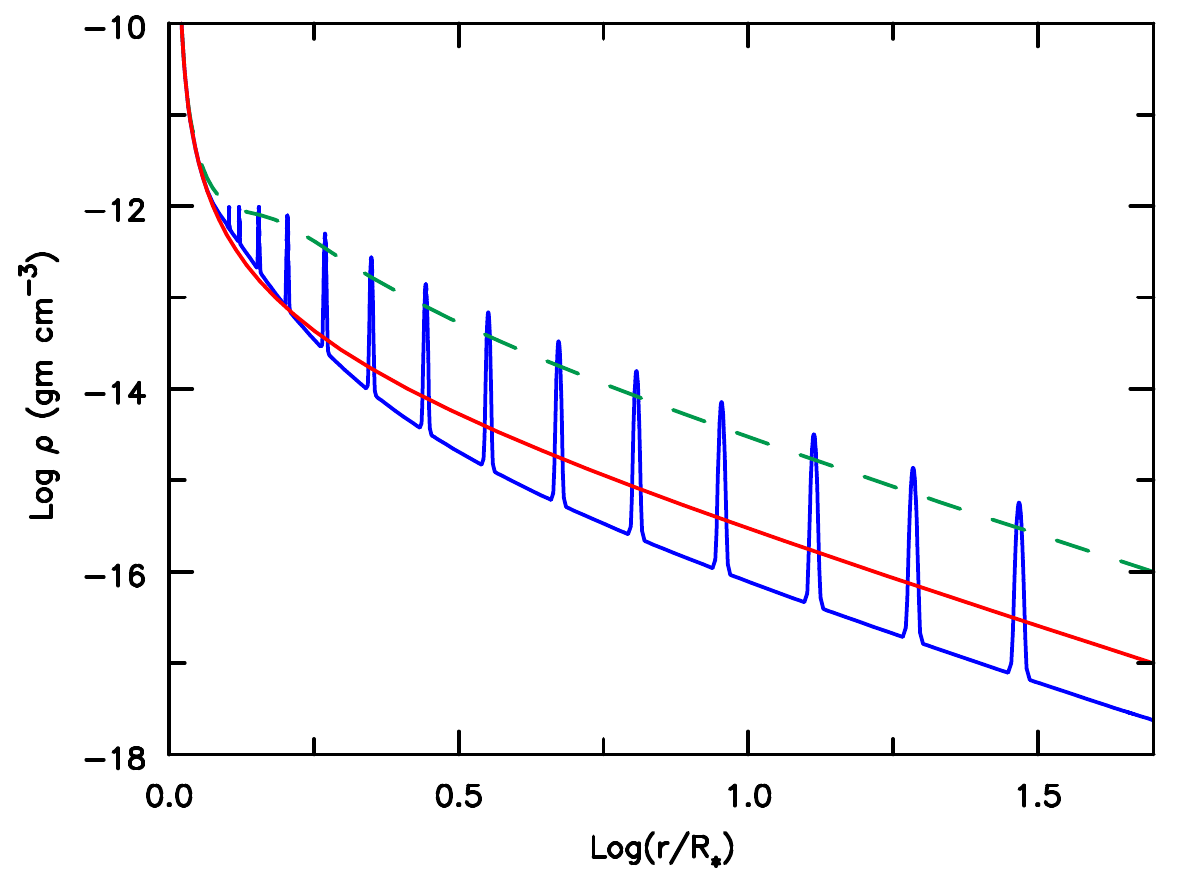}
	\caption{Comparison of the density distribution in the wind. The red solid curve is the wind's smooth density distribution, the blue solid curve is the Shell model, and the green dash curve corresponds to the VFF model with $f_\mathrm{v}=0.1$. }
	\label{fig_dens}
\end{figure}

\begin{figure}  
	\includegraphics[width=0.95\columnwidth]{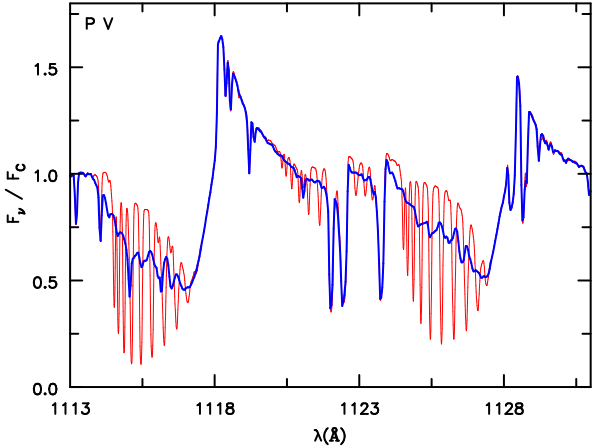}
		\caption{Illustration of the \ion{P}{V} $\lambda\lambda$1118 and 1128 profiles that were synthesized using \textsc{cmf\_flux} (\textit{red, thin curve}) and \textsc{obs\_frame\_2d} (\textit{blue, thick curve}) using the same Shell model. Removing the influence of rotation and fixing the microturbulent velocity to 10\,\kms, the  \textsc{cmf\_flux} spectrum shows a complex absorption structure that arises from the use of spherical shells. This structure is related to velocity porosity, and is removed by assuming that shells along different sight lines have a slightly different radial velocity.}
	\label{fig_shell_art}
\end{figure}

The VFF approach uses an ad\~hoc, simple parametric treatment motivated by hydrodynamical simulations. It assumes the wind is comprised of uniformly distributed, optically thin clumps with a void ICM. In \textsc{cmfgen}, the wind is assumed to be homogeneous close to the star, then becomes clumpy at a characteristic velocity scale $V_\mathrm{cl}$. It reaches full ``clumpiness'', with a volume-filling factor $f_\mathrm{v}$, at large radii. Mathematically, 
\begin{equation}
	f(r)= f_\mathrm{v} + (1-f_\mathrm{v})\exp(-V(r)/V_\mathrm{cl})\,\,.    
	\label{eq:fcl}
\end{equation}

In the Shell approach, we explicitly insert ``shells'' into the wind -- highly dense regions equally spaced logarithmically, scaled to the star's radius. Several parameters are introduced to control the profile of the shells and ICM while assuming the local mean density is proportional to the local smooth wind density.

The Shell model begins with the construction of a shell profile; the profile is modeled with a Gaussian distribution function which is mainly controlled by its Full-Width Half-Maximum (FWHM) value and the characteristic radial distance scale R\textsubscript{scl}. The number of depth points along the shell, ND\textsubscript{shell}, is increased and adjusted such that the shell is resolved. Figure~\ref{fig_shell_prof} illustrates a shell profile used in the Shell model presented in this paper. The values for FWHM and R\textsubscript{scl} are chosen such that the calculated effective volume-filling factor is comparable to the adopted volume-filling factor value in the VFF model. After the shell profile is created, we insert multiple copies into a homogeneous wind (red curve in Fig. \ref{fig_dens}). Below the velocity V\textsubscript{onset}, the wind is assumed to be homogeneous. Above V\textsubscript{onset} shells are inserted at a characteristic radial distance scale, R\textsubscript{scl}, while the ICM is scaled with the characteristic density contrast scale, V\textsubscript{clump}. The resulting density distribution is the blue solid curve in Fig. \ref{fig_dens}.  After comparing spectra of several Shell models to observations, the values listed in  Table \ref{table:Param} are the model parameters that provide the best fit to the spectra of AzV83.

After the \textsc{cmfgen} model has converged, we compute the observed spectrum using the comoving/observer's frame code \textsc{cmf\_flux}. This code first evaluates all emissivities and opacities in the comoving frame, and then converts them into the observer's frame for an observer's frame calculation. In these calculations, we account for electron scattering, Doppler effects (higher order relativistic terms are not necessary for O stars) and microturbulence. We model the microturbulent velocity field using the standard assumption of Gaussian microturbulence and assume that it is depth-dependent and can be modeled by the following linear function: 
	\begin{equation}
		\varv_\mathrm{turb}(r)=\varv_\mathrm{min} + ( \varv_\mathrm{max}- \varv_\mathrm{min})
                 \times \dfrac{V(r)}{V_\mathrm{\infty}}.
		\label{eq:vturb_law}
	\end{equation}
In this expression $\varv_\mathrm{min}$ is the minimum turbulent velocity in the photosphere, and $\varv_\mathrm{max}$ is the maximum turbulent velocity in the wind.

For the VFF model, the spectrum produced by \textsc{cmf\_flux} is generally adequate for comparing with observation\footnote{As noted in the text we use convolution to allow for the influence of rotation for synthesized spectra using the VFF method. This works well for the majority of photospheric features but can fail for features affected by emission \citep{Hillier2012}. The 2D code, \textsc{obs\_frame\_2d}, provides more realistic line profiles for these cases.}. However, for the Shell model we use the 2D code, \textsc{obs\_frame\_2d} \citep{Busche2005}. \textsc{obs\_frame\_2d} takes the opacities, emissivities and atmospheric structure from \textsc{cmf\_flux} and \textsc{cmfgen} and maps these into 3D. The observed spectrum is then constructed by computing the emergent intensity along a set of rays that are specified by two polar co-ordinates --- the impact parameter and  azimuth ( $\varphi$). A 2D integration of these intensities is then performed to compute the final spectrum.

When mapping the emissivity and opacities into the observer's frame, the velocity along each ray is randomly scaled (e.g., $V(r)_{\rm scaled} =V(r)(1+w dV)$ where $dV$ is the maximum supplied scaling (typically 0.2) and $w$ is a random number between -1 and 1. This introduces a small random shift in the velocity of each shell which smooths artefacts in the P~Cygni profiles that arise from the discrete and spatially coherent spherical shells. Figure~\ref{fig_shell_art} illustrates both the artefacts and the smoothing effect -- the random scaling only influences the absorption component of the P~Cygni profile, and does not influence the emission component.  While the use of the 2D code does introduce an inconsistency, the effects are likely to be small for the present models since the absorption of continuum radiation along a given ray is independent of what occurs for other rays striking the core, and is independent of other shells. 
Spectra computed using the 1D code, and those computed using the 2D code, show good agreement apart from the smoothed absorption  components. As apparent from Fig.~\ref{fig_shell_art}, the 2D code yields an absorption structure that would, to a good approximation, occur from smoothing the absorption component of the 1D code.

Broadening due to rotation is done explicitly for spectra synthesized with \textsc{obs\_frame\_2d} while spectra from \textsc{cmf\_flux} have to be convolved with a rotational-broadening function. However, to ensure consistency, spectra for both the VFF and Shell models have been synthesized using \textsc{obs\_frame\_2d}. Based on earlier modeling of AzV83 by \cite{Hillier2003}, we adopt a rotational velocity $\varv \sin i = 70\,\kms$. 

%%%%%%%%%%%%%%%%%%%%%%%%%%%%%%%%%%%%%%%%%%%%%%%%%%%%%%%%%%%%%%%%%%%%%
%%%%%%%%%%%%%%%%%%%%%%%%%%%%%%%%%%%%%%%%%%%%%%%%%%%%%%%%%%%%%%%%%%%%%

 \begin{figure}%ionisation structure
 \includegraphics[height=17.5cm]{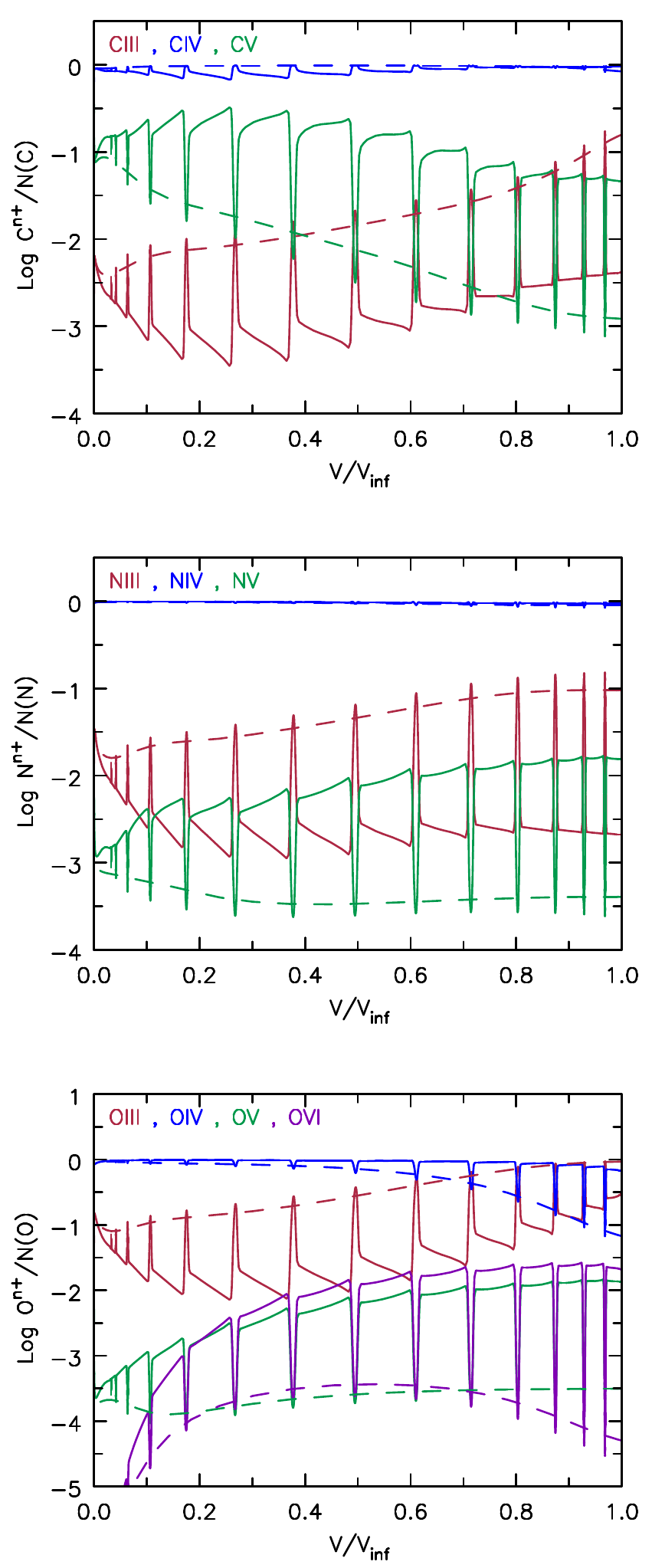}
	\caption{Ionisation structure of a few species in the winds of  a Shell \textit{(solid curves)} and a VFF \textit{(dashed curves)} model as a function of normalized velocity. The ionisation structure of the VFF model varies smoothly with increasing velocity while the Shell model's does not -- the ionisation fraction in the ICM may differ by as much as three orders of magnitude from the neighbouring shell. For example, \ion{O}{III} has a higher fraction within the dense shell compared to its surrounding ICM, whereas the opposite is true for \ion{O}{IV}, \ion{O}{V}, and \ion{O}{VI}.}
	\label{fig_IF}
\end{figure} 

\subsection{X-rays}

O stars are X-ray emitters and there is strong evidence (for single stars) that X-rays arise from shocks in the
stellar wind \citep[e.g.,][]{1983ApJ...271..681C,2014MNRAS.439..908C}.  Observations show that the X-ray luminosity is strongly correlated to the luminosity with $\log (L_\mathrm{x} / L_*) \approx -7 \pm 1 $  \citep{Pallavicini1981,Chlebowski1989}. \textsc{cmfgen} does not compute the X-ray luminosity from first principles. Instead several control parameters are used -- these specify the location at which the shocks begin, and a characteristic shock velocity. For the X-ray emissivity we use a table of X-ray spectra computed with the Raymond-Smith plasma code (other choices are also available). 

While the X-ray luminosity  is small it can influence the ionisation structure via direct photoionisation, and 
 via the Auger ionisation mechanism \citep{Cassinelli1979}. In the latter mechanism, a high-energy X-ray ejects an inner shell electron (a 1s electron for CNO). Immediately afterward, the highly excited ion will reconfigure itself, resulting in a second electron being ejected. This mechanism has been widely accepted as the cause of super-ion lines -- P~Cygni lines (e.g., \osixdoub)\ arising from species that cannot be produced by the photospheric radiation field.  
 
\section{Results}\label{sec_results}

\subsection{Ionisation Structure}\label{subsec_ion_temp} 

\begin{figure}
	\includegraphics[width=0.95\columnwidth]{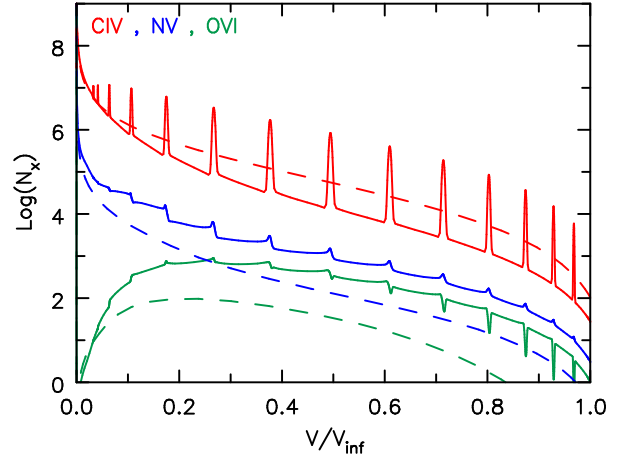}
	\caption{Number population of \ion{C}{IV}, \ion{N}{V}, and \ion{O}{VI} -- shown in red, blue, 
	and green, respectively -- for the Shell \textit{(solid curves)} and VFF \textit{(dashed curves)} model. All models have the same X-ray luminosity ($\log (L_\mathrm{X} / L_*) = -7 $). In the VFF model, \ion{N}{V} and \ion{O}{VI} are confined to the clumps -- there is
 no interclump medium. On the other hand, in the Shell model, the density of these species within the ICM is comparable to that in the shell. Thus, the ICM plays a significant role in the ionisation balance and spectral profile of \ion{N}{V} and \ion{O}{VI}. The number density of \ion{C}{IV} does not show this characteristic and is dependent on the density of the wind.} 
	\label{fig_superion_pop}
\end{figure}

Incorporating shells into the stellar wind model allows us to study the influence of the dense shells  and ICM on the ionisation structure of the wind. The series of plots in Fig. \ref{fig_IF} compares the ionisation structure of the Shell model with the VFF model as a function of normalized velocity. Qualitatively, the ionisation structure of the Shell model follows the general trend of the VFF ionisation structure -- higher ionized states decrease while lower ionized states increase at larger velocities.  However, in the Shell model the ionisation does not vary smoothly with radius -- the ionisation fraction in the ICM differs from that in the core of the shell by as much as three orders of magnitude. Referring back to Fig. \ref{fig_dens}, these differences are explained by the contrasting density between the ICM and the shells themselves -- recombination rates are higher within the shells than in the ICM. 

With the inclusion of X-rays, the consequence of using the Shell approach becomes apparent when studying the relative importance of the ICM on the super-ions' number density. Figure \ref{fig_superion_pop} plots the populations of \ion{C}{IV}, \ion{N}{V}, and \ion{O}{VI} as a function of normalized velocity. The number densities of \ion{N}{V} and \ion{O}{VI} in the Shell model vary relatively smoothly with velocity. Further, the number densities are significantly higher than in the VFF models. This clearly illustrates the importance a non-void ICM has on the ionisation structure of super-ions and, by inference, on their UV resonance profiles (Section~\ref{subsec_morph}). The role of the ICM in the formation of the super-ion profiles is as crucial as the clumps themselves but the VFF method fails to take this into account. The reason for the importance of the ICM for the super-ion profiles is extensively discussed by \cite{Zsargo2008}.

In AzV83, \ion{C}{IV} is not a super-ion, and it behaves differently from both \ion{N}{V} and \ion{O}{VI}. This is easy to explain --  \ion{C}{IV} is the dominant ionisation stage, whereas \ion{N}{IV} and \ion{O}{IV}  are the dominant ionisation stages for N and O. 

In addition to differences in ionization between the VFF and Shell models, there are also differences in level populations. We illustrate the latter in Appendix~\ref{Non-LTE Effects} for the $n=2$ and $n=3$ levels of hydrogen.  The differences are complicated but are expected. They arise from the asymmetry of the radiation field, the finite size of the shells, and the higher and variable density of the shells. Since the spectra obtained using the Shell or VFF approach are similar (similar), and since AzV83 has limited wind diagnostics we will not address the differences here. However, they will be addressed in a forthcoming paper where we study a Wolf-Rayet star of type WNE, and where we find large spectral differences between the VFF and Shell approaches.
 
%%%%%%%%%%%%%%%%%%%%%%%%%%%%%%%%%%%%%%%%%%%%%%%%

\subsection{Spectral Comparison} \label{subsec_morph}

\begin{figure*}
	\includegraphics[scale=0.85]{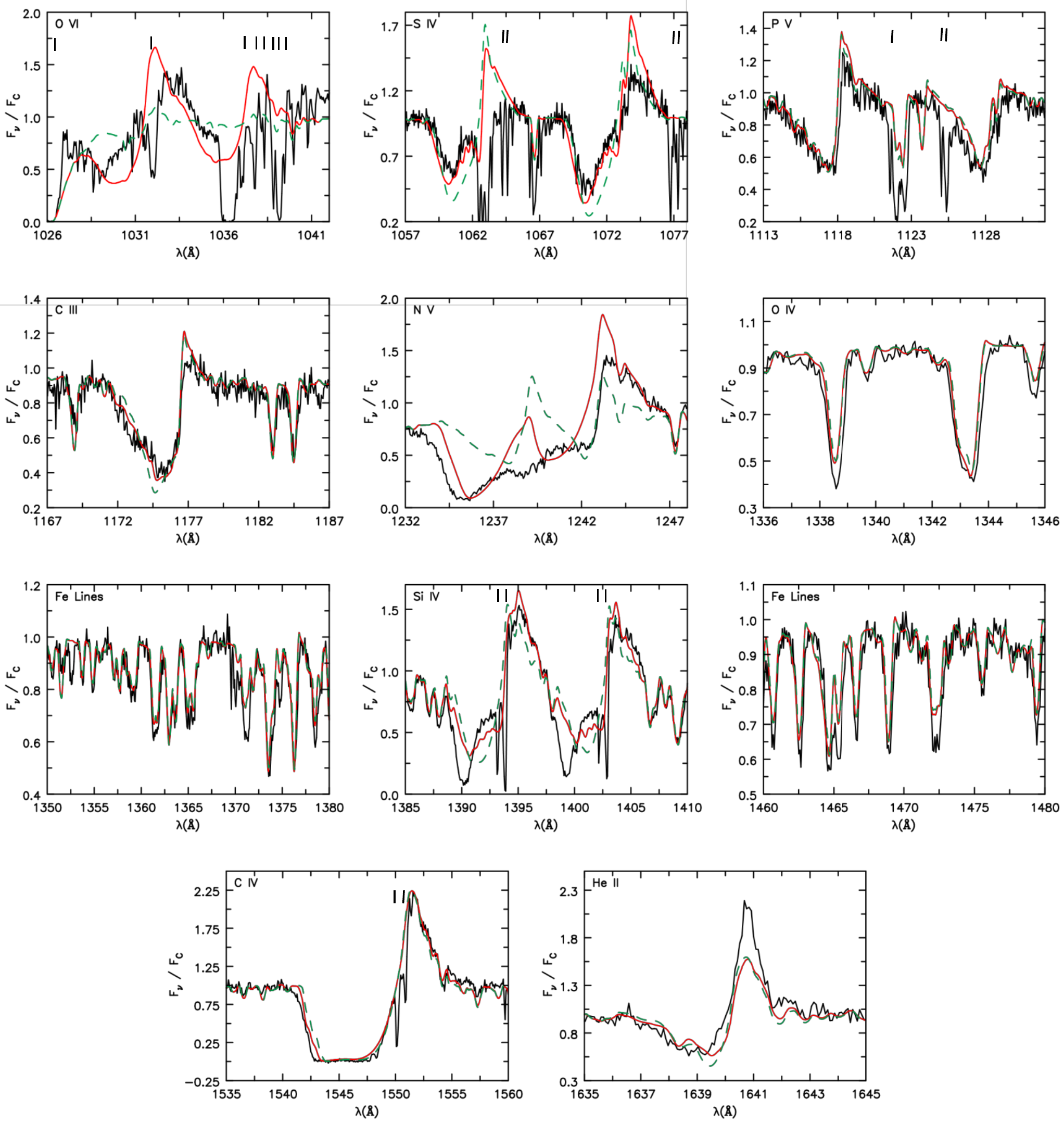}
	\caption{UV spectra of AzV83 \textit{(black rough curve)} compared with \textsc{cmfgen} models using VFF approach \textit{(green dashed curve)} and Shell approach \textit{(red solid curve)}. The black tick-marks indicate the presence of interstellar lines seen in the observed spectra that we do not model. Note that the Shell model had its mass-loss rate and terminal velocity revised, from $8.0 \times 10^{-7}$ to $1.1 \times 10^{-6} \,\Msunyr$ and from 862 to 1000\,\kms\ respectively, to better fit the breadth of the absorption component of the resonance lines. Both approaches result in similar iron lines but differ in the profiles of the resonance lines. Significant improvement in fitting the super-ions -- the \ion{O}{VI} $\lambda$$\lambda$1031,1037 and \ion{N}{V}  $\lambda$$\lambda$1238,1242 doublets -- is achieved when using the Shell approach; this, again, stems from the inclusion of an ICM in the wind. 
	}
	\label{fig_UV}
\end{figure*}

\begin{figure*}
	\begin{minipage}[t]{\linewidth}
		\centering
		\includegraphics[scale=0.85]{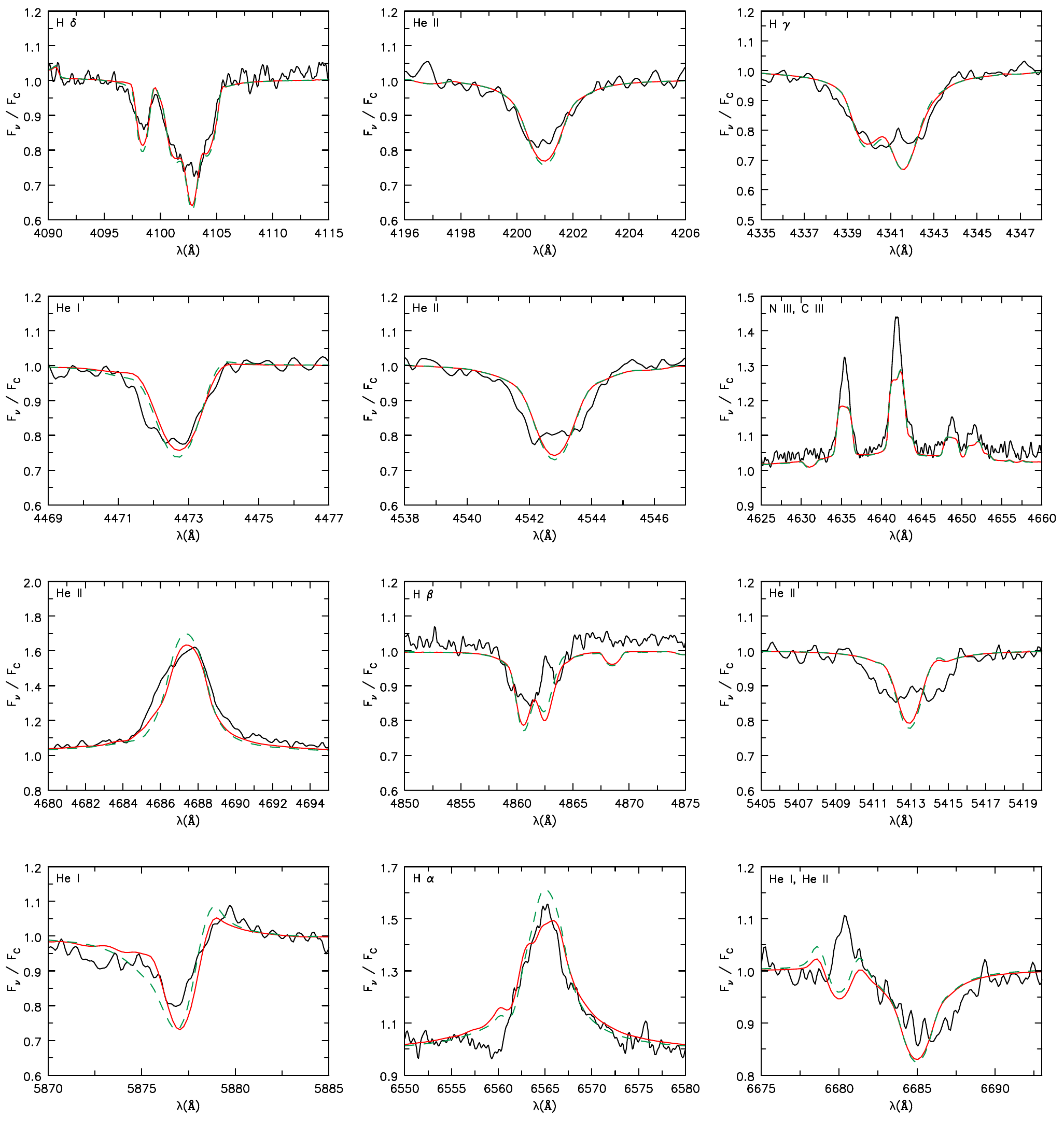}
	\end{minipage}
	\caption{Optical spectra of AzV83 \textit{(black rough curve)} compared with \textsc{cmfgen} models using VFF approach \textit{(green dashed curve)} and Shell approach \textit{(red solid curve)}.   Note that the Shell model had its mass-loss rate and terminal velocity revised, from $8.0 \times 10^{-7}$ to $1.1 \times 10^{-6}$ \Msunyr\ and from 862 to 1000\,\kms\ respectively, to improve the fit to the \ion{He}{II} $\lambda$4686 and H$\alpha$ emission. }
	\label{fig_OPT}
\end{figure*}

Several prominent features in the UV and optical spectra of AzV83 are plotted in Figs. \ref{fig_UV} and \ref{fig_OPT}, and compared to spectra synthesized from \textsc{cmfgen} models that used either the VFF or Shell approach. In general the synthesized spectra from the two approaches are similar, and produce similar qualitative agreement with the observations. The largest changes occur for the profiles of  \ion{N}{V} and \ion{O}{VI} -- the two lines most affected by both the ICM and X-rays.

To better fit the breadth of the absorption component of the resonance lines and the line strengths of \ion{He}{II} $\lambda$4686 and H$\alpha$, the Shell models had their mass-loss rate and terminal velocity revised, from $8.0 \times 10^{-7}$ to $1.1 \times 10^{-6} \,\Msunyr$ and from 862 to 1000\,\kms\ respectively.
The weakening of the lines can be explained by optical depth effects in the clumps and by a reduction in the effective recombination rate into the upper levels. The latter arises from the lower clump density as the photosphere is approached. Thus, in order to compensate for a lower recombination rate, a larger mass-loss rate was used. This effect was also observed by \cite{Sundqvist2011}. 

In subsequent subsections we analyze specific features seen in  UV and optical spectra, and discuss how their behaviour is dependent on which of the two methods is used for treating clumped winds.

\subsubsection{\ion{N}{V} and \ion{O}{VI} Resonance Lines and X-ray Luminosity} \label{subsec_xrays}

\begin{figure*} 
	\begin{minipage}[t]{\linewidth}
		\centering
		\includegraphics[width=14cm]{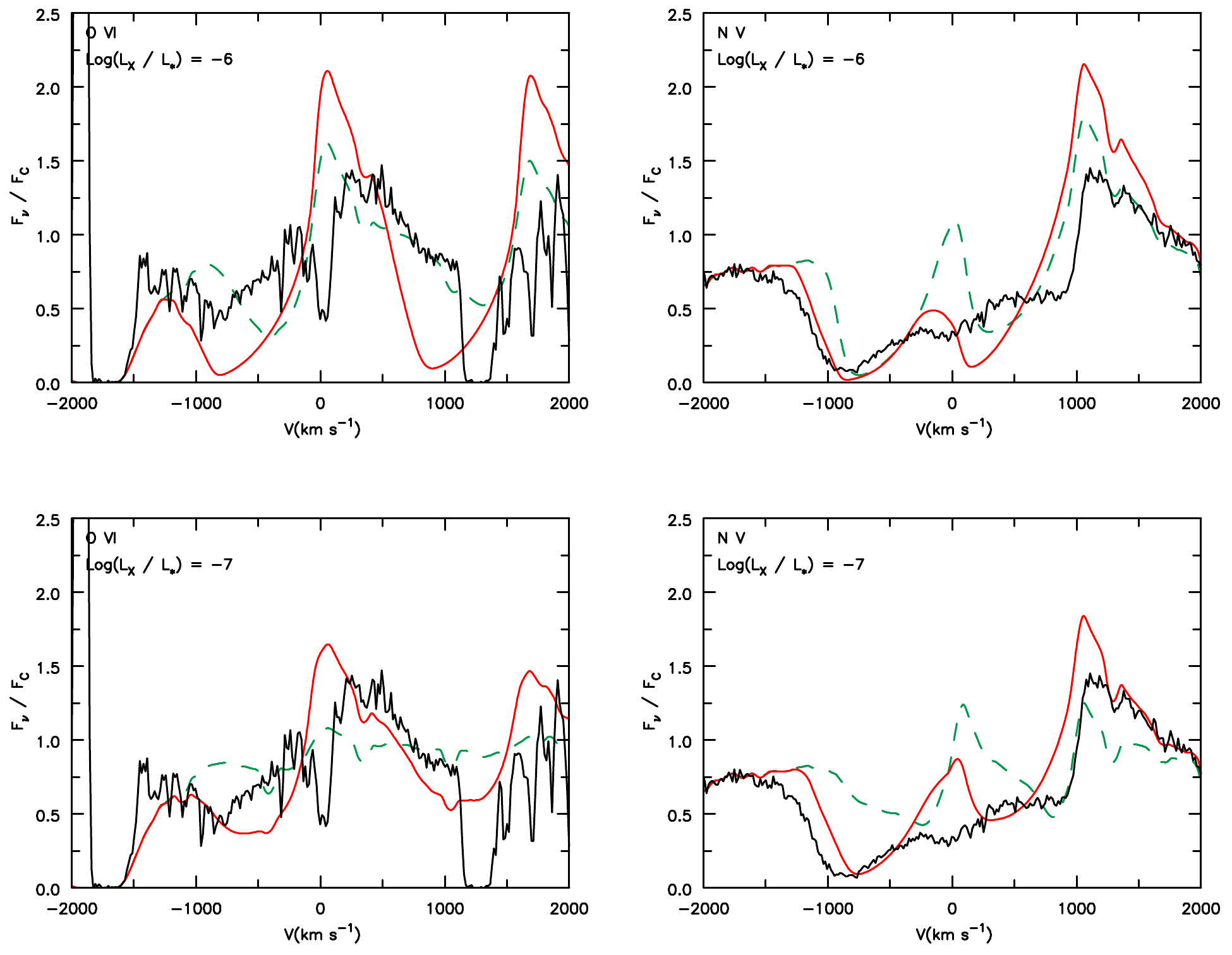}
	\end{minipage}
	\caption{Plots of \ion{O}{VI} $\lambda$1031 and \ion{N}{V} $\lambda$1238 for both the Shell \textit{(red solid curve)} and VFF \textit{(green dashed curve)} models in velocity space for a maximum turbulent velocity of $\varv_\mathrm{turb}$= 100 \kms. Two values of X-ray luminosity are plotted here: Log$(L_\mathrm{x} / L_*) = -6 $ \textit{(top row)} and $-7$ \textit{(bottom row)}. The rough, black curve is the observed spectrum.   Spectra synthesized from the VFF model fail to fit the observed UV resonance lines of \ion{O}{VI} and \ion{N}{V} due to its inadequate assumption of a void ICM. The VFF model requires an unusually high X-ray emission in order to fit the synthesized spectra to observations. No adjustment of the stellar parameters or wind velocity law was made to improve the Shell model fit to the observed lines.
	}
	\label{fig_LX}
\end{figure*}

As mentioned earlier  (Section~\ref{subsec_ion_temp}), the resonance line profiles of \ion{N}{V}\ and \ion{O}{VI}\ depend not only on the density structure of the wind, but also on the X-ray luminosity.  In Fig. \ref{fig_LX}, we have plotted several model spectra for the \ion{O}{VI}  $\lambda$$\lambda$1031,1037 and \ion{N}{V} $\lambda$$\lambda$1238,1242 doublets using the VFF and Shell approaches, for two values of the X-ray luminosity. For an X-ray luminosity of Log$(L_\mathrm{x} / L_*) = -7$, we immediately see that the VFF model fails to fit the observed spectra, whereas the Shell model better reproduces the observed  \ion{N}{V} and \ion{O}{VI} doublet profiles. Additionally, Fig. \ref{fig_LX} illustrates that in order for the VFF model to reproduce these profiles,  the X-ray luminosity would need to be an order of magnitude higher.  

This result is not unexpected. Models using the VFF approach have trouble fitting
these lines due to the high density in the clumps, which enhances recombination, and severely curtails the abundance of \ion{O}{VI} and \ion{N}{V} for a given X-ray flux. On the other hand, the Shell model contains low density regions in which recombination is relatively inefficient. As discussed by \cite{Zsargo2008}, the super-ion population in the ICM can be similar to that in the clumps. The ICM can also play a role in compensating for the effects of porosity in velocity space  \citep{Sundqvist2011}.

\subsubsection{UV Resonance Lines}\label{subsec_res_lines}

\begin{figure}  
	\begin{minipage}[t]{\linewidth}
		\begin{subfigure}{.5\textwidth}
			\centering
			\includegraphics[scale=0.63]{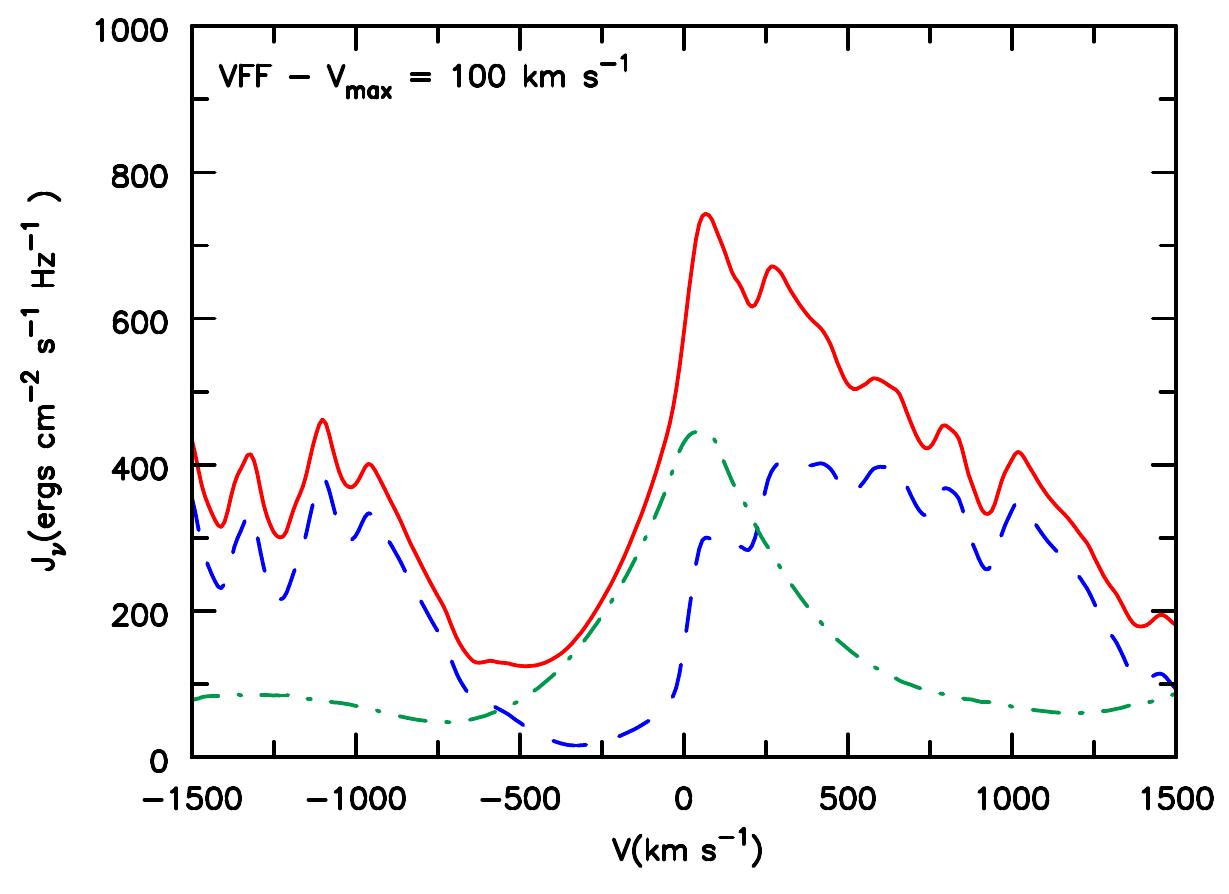}
		\end{subfigure}%
		\vskip 0.25cm
		\begin{subfigure}{.5\textwidth}
			\centering
			\includegraphics[scale=0.80]{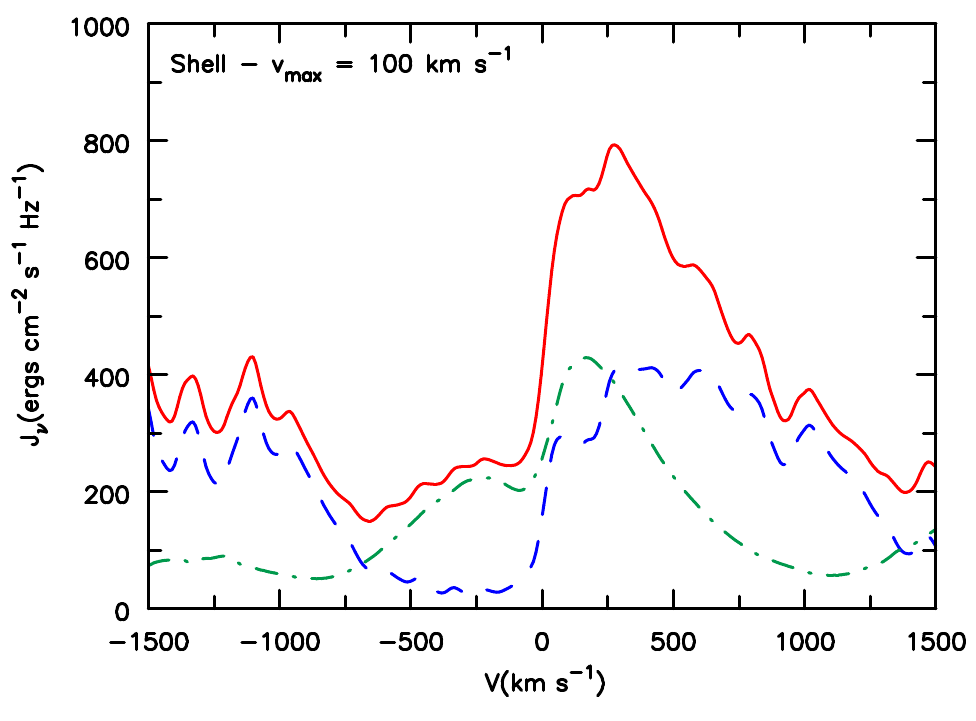}
		\end{subfigure}%
	\end{minipage}
	\caption{Plots of the flux of the \ion{Si}{IV} $\lambda$1393 resonance transition, integrated over  AzV83's stellar disk \textit{(blue, dashed curve)} and outside the disk \textit{(green, dot-dash curve)} for the VFF \textit{(top)} and Shell \textit{(bottom)} models. The red, solid curve is the sum of the two other curves and represents the modeled spectra we would obtain from \texttt{CMF\_FLUX} and \texttt{OBS\_FRAME\_2D}; no scaling or rectification has been performed. The figure illustrates how the formation of the scattering component of the P Cygni profile hinges upon which method is implemented for treating clumping.
	}
	\label{fig_IP}
\end{figure}

\begin{figure*}  
	\begin{minipage}[t]{\linewidth}
		\centering
		\begin{subfigure}{.5\textwidth}
			\centering
			\includegraphics[width=8cm]{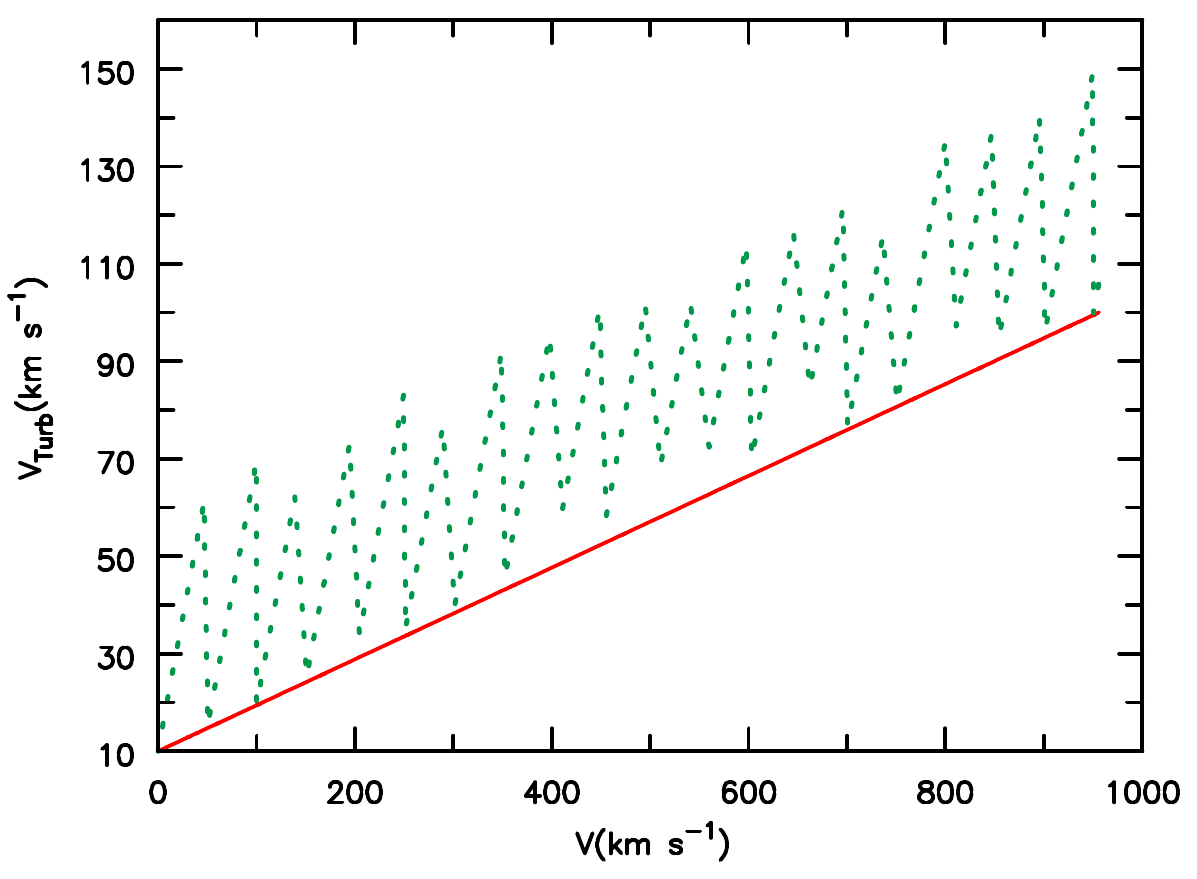}
		\end{subfigure}%
		\begin{subfigure}{.5\textwidth}
			\centering
			\includegraphics[width=8cm]{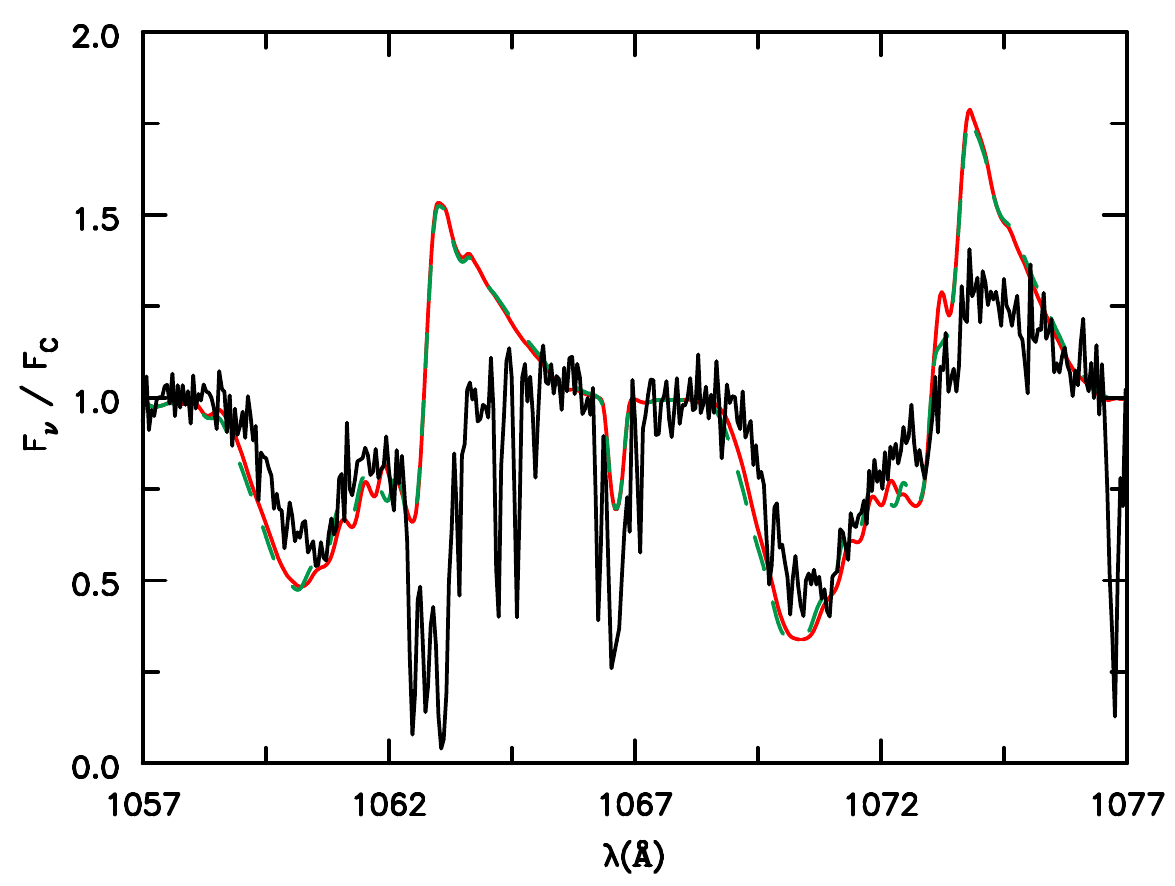}
		\end{subfigure}%
		\vskip 0.25cm	
		\begin{subfigure}{.5\textwidth}%
			\centering
			\includegraphics[width=8cm]{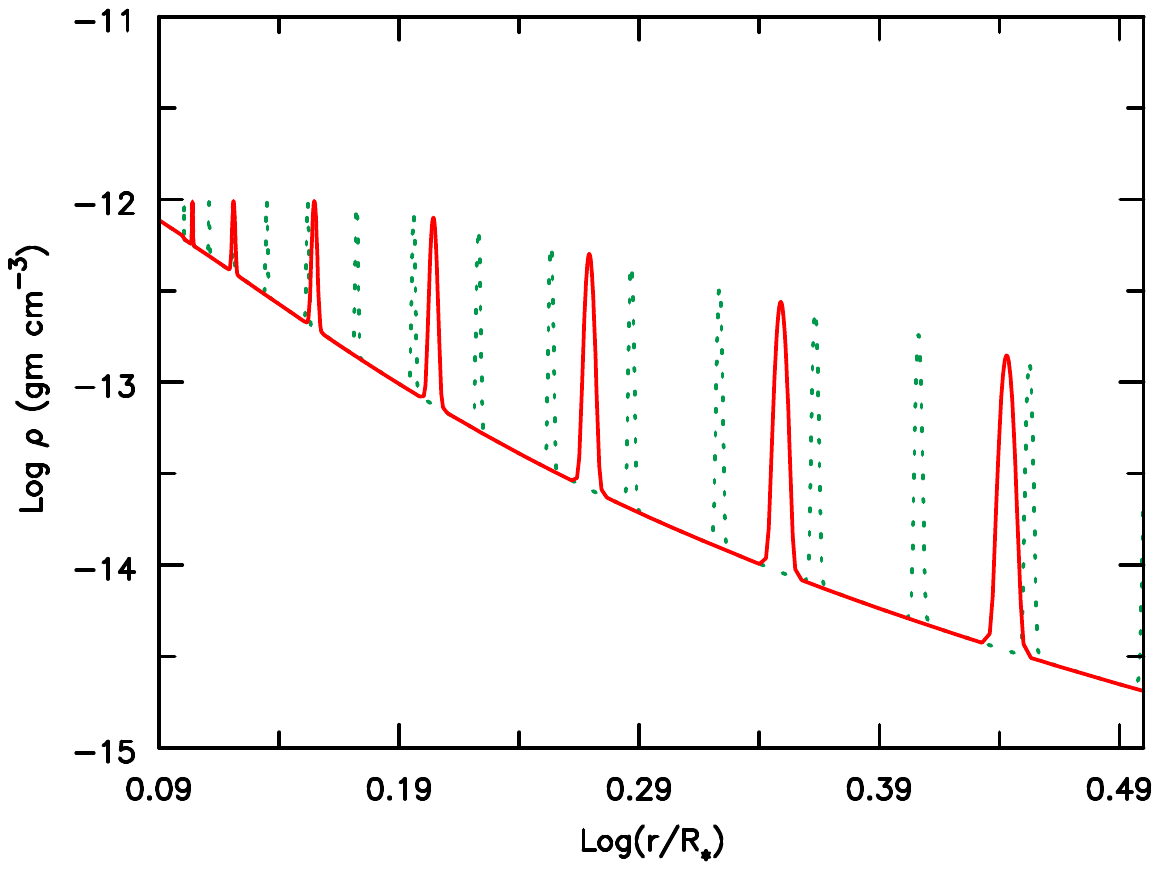}
		\end{subfigure}%
		\begin{subfigure}{.5\textwidth}%
			\centering
			\includegraphics[width=8cm]{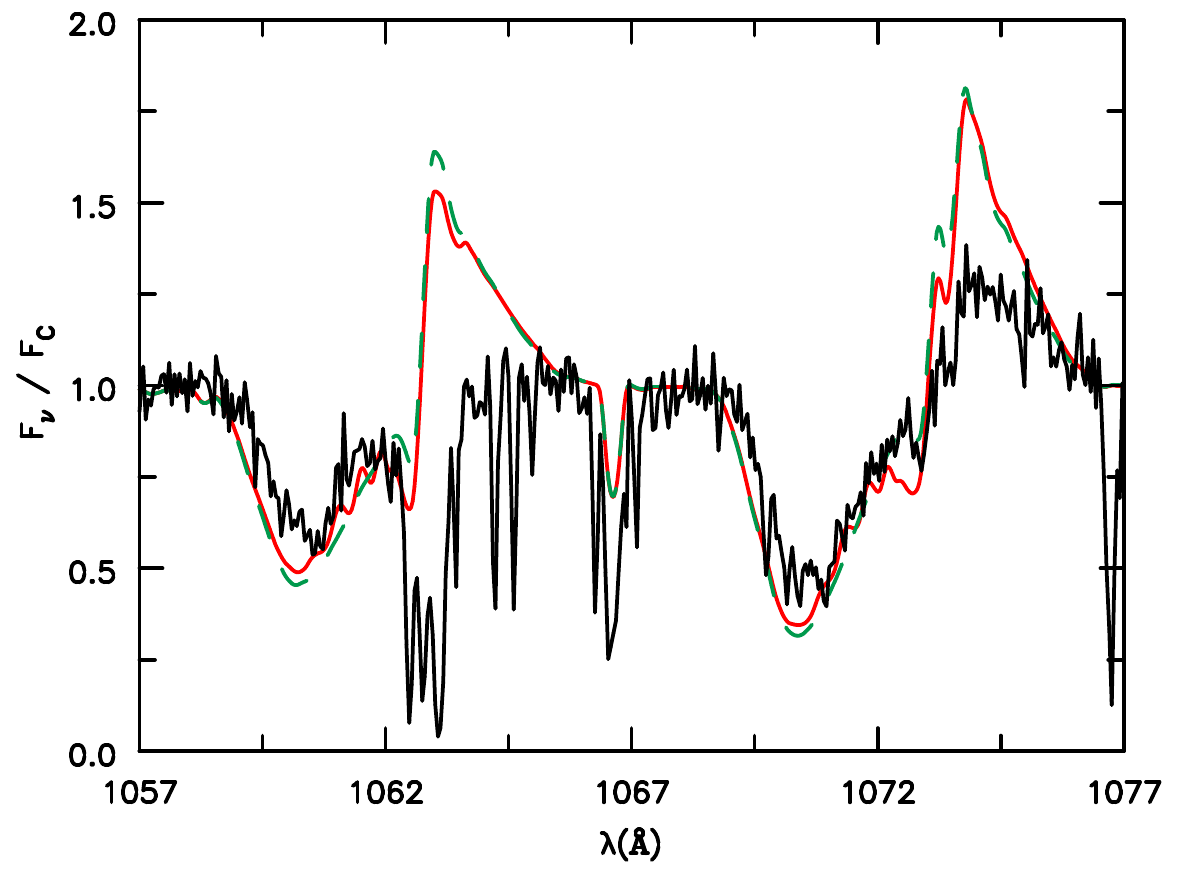}
		\end{subfigure}%
	\end{minipage}
	\caption{We plot two sets of experiments (\textit{green, dotted curves}) in an attempt to correct for the excess red-emission of \ion{S}{IV} $\lambda\lambda 1062 , 1073$ (\textit{red curves}). We experimented by modifying our treatment of microturbulent velocity (\textit{top left}) by superimposing a seesaw-like velocity field; the resulting spectra (\textit{top right}) shows a small shift in its blue trough but retains high red emission. We also explored increasing the number of shells in the wind (\textit{bottom left}); the resulting spectra (\textit{bottom right}) are virtually identical to each other. 
	}
	\label{fig_nonmonvel}
\end{figure*}

Although the Shell approach reproduces the general P Cygni profiles as in the VFF models (see Fig. \ref{fig_UV}) their formation is 
different; the emission component of the P Cygni profile, formed by resonance scattering, is strongly tied to the matter distribution in the wind, and hence the method chosen  to treat clumping. To illustrate this, Fig. \ref{fig_IP} plots the flux of \ion{Si}{IV} $\lambda 1393$, integrated over the cross-section of AzV83's stellar disk (\textit{blue-dash curve}) and its wind (\textit{red-solid curve}) for both approaches. The line,  integrated over the stellar disk,  exhibits blue-shifted absorption, and is very similar for both clumping approaches. This contrasts with the emission component arising from the wind -- \textsc{cmfgen} models that use the VFF approach to treat clumping have a symmetric emission profile about line centre, whereas models that use the Shell approach have an asymmetric profile, favouring redward emission.

The reason why we see this in our spectra is as follows. In the VFF approach the wind is assumed to vary smoothly -- homogeneous close to the star, becoming progressively clumpy at larger radii. As such, every point along the wind is a site for emission, leading to a symmetric emission profile as seen in the top panel of Fig. \ref{fig_IP}. However by redistributing the wind material into shells, emission will occur solely in the shells, with none in the ICM. No shell-like structure appears in the emission profile as a result of using \textsc{obs\_frame\_2d} where shells along different sight lines have a slightly different radial velocity. The dip in the emission profile at line centre is an optical depth effect of using shells as our clumps; photons emitted at line centre have to traverse through a larger cross-section of the shell, reducing the flux. 

However, this does not explain why both sets of models show excess emission on the red side of the line profiles that is not observed  -- see \ion{S}{IV} $\lambda$$\lambda$1062,1073; \ion{P}{V} $\lambda$$\lambda$1117,1128; \ion{C}{III} $\lambda$1177; and \ion{Si}{IV} $\lambda$$\lambda$1393, 1402 in Fig. \ref{fig_UV}. We performed several experiments to rectify this excess, with the results shown in Fig.~\ref{fig_nonmonvel}. We toyed with our microturbulent velocity by superimposing a seesaw-like velocity field in the \textsc{obs\_flux} code; the resulting velocity field is qualitatively similar to that used in \cite{Lucy1982}. This experiment explored whether the non-monotonic velocity field influences the shells' ability to efficiently destroy these excess line photons. While we found that the absorption trough deepened slightly, the emission peak remained similar. In another test we created a Shell model with more shells while maintaining a similar effective volume-filling factor. This was to test whether having more shells could extend the location in the wind where optical depth unity occurs, effectively reducing the emergent flux. However, here too we see insignificant differences in the spectra.

 While the Shell model yielded similar fits to the VFF models for the observed UV resonance lines, the consistent excess emission from both methods is concerning. This is a systemic error that needs to be addressed in future modeling -- it is an indicator that we are not capturing the correct physics in the wind or that we are missing a key ingredient in the modeling. Further investigations into this discrepancy are required.

\subsubsection{Microturbulence}\label{sub_micro}

We followed up on several lines that \cite{Hillier2003} found to have some dependence on the adopted microturbulent velocity: the \ion{Fe}{V} spectrum in the UV, the \ion{O}{IV} triplet $\lambda\lambda 1339 -1344$, and \ion{He}{II} $\lambda$5413. As noted in Section \ref{sec_obs_mod}, we adopted the simple depth-dependent turbulence formula, with 10 \kms and 100 \kms\ as our microturbulent velocity at the base of the wind and at large radii, respectively. 

\begin{figure*}
	\begin{minipage}[t]{\linewidth}
		\begin{subfigure}{1\textwidth}
			\centering
			\includegraphics[width=1\linewidth]{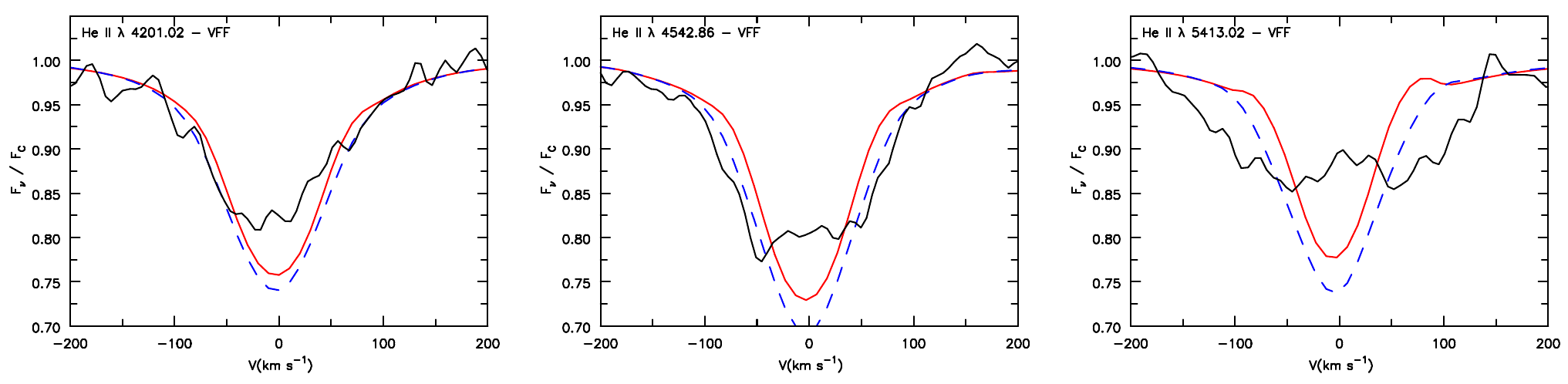}
		\end{subfigure}
	\end{minipage}	
	\caption{Illustration of the \ion{He}{II} $\lambda$4201, $\lambda$4542, and $\lambda$5413 line profiles and their dependence on the minimum microturbulent velocity for the VFF model: $\varv_\mathrm{min}$= 10 \kms \textit{(red, solid curve)} and  $\varv_\mathrm{min}$= 25 \kms \textit{(blue, dashed curve)}. The plots for the Shell model are not shown as they are very similar to those of the VFF model.   As you move redward, from \ion{He}{II} $\lambda$4201 to $\lambda$5413, the dependence on microturbulence becomes stronger across both models as a result of where their line formation region lies -- \ion{He}{II} $\lambda$4201 and $\lambda$4542 lines form just below the sonic point and \ion{He}{II} $\lambda$5413 above it. However, our models could not capture the unusually broad profile of \ion{He}{II} $\lambda$5413. 
	}
	\label{fig_HeII_turb}
\end{figure*}

While our best-fit Shell model was able to fit the observed \ion{Fe}{V} spectrum and \ion{O}{IV} triplet $\lambda\lambda1339 -1344$, the fit to \ion{He}{II} $\lambda$5413 worsened -- \ion{He}{II} $\lambda$5413 became narrower although its equivalent width remained similar to that of the VFF model\footnote{Towards the end of our analysis, we discovered that our models used an old hydrostatic structure computed with less sophisticated models. When the hydrostatic structure was updated, some minor variations did occur, including a change in \ion{He}{II} $\lambda$5413 -- about a 5\% difference in equivalent width. However, correcting the hydrostatic model did not improve our fit for \ion{He}{II} $\lambda$5413 -- our model still predicts a narrower line compared to observation. As such, our analysis described in this subsection remains unchanged.}.  None of our models are able to fit the \ion{He}{II} $\lambda$5413 profile which is unusually broad  -- this characteristic is also seen in another observational dataset supplied by Philip Massey (private communication). To further investigate this line, we examined two other isolated members of the Pickering series -- \ion{He}{II} $\lambda$4201 (n = 11 to 4) and  $\lambda$4543 (n = 9 to 4).  For both sets of models, the fits to the Pickering series are better for higher series members (Fig.~\ref{fig_OPT}). As illustrated in Fig. \ref{fig_HeII_turb} the higher series members are narrower than \ion{He}{II}\ $\lambda$5413, and are less sensitive to the microturbulent velocity.

This dependence on microturbulent velocity arises because these lines form in different regions; \ion{He}{II} $\lambda$4201 and $\lambda$4543 form just below the sonic point of the wind whereas \ion{He}{II} $\lambda$5413 forms just above it. Given the morphology of \ion{He}{II} $\lambda$5413, compared to \ion{He}{II} $\lambda$4201 and $\lambda$4543, this may suggest that higher turbulence is occurring just above the sonic point. The source of this enhanced turbulence is unknown, but may be a result of ``failed winds'' falling back onto AzV83. In any case, this led us to reevaluate how we treat microturbulence. 

In an attempt to rectify the strength of \ion{He}{II} $\lambda$5413, we began by moving away from modeling depth-dependent turbulence as a linear function, as shown in Eq. \ref{eq:vturb_law}. Several models were considered and we decided to model turbulence as a spliced function that for this paper we call the \textit{Linear-CubeRoot-Linear} (LCL) function. It begins by describing turbulence as a linear function as in Eq. \ref{eq:vturb_law}, then accelerates as a cube-root function, and ends linearly again. By situating the function's acceleration portion close to where $\log \tau$(\ion{He}{II} $\lambda 5413$) = 1.0, it could describe the situation where significant turbulence is occurring at that region. $\tau$(\ion{He}{II} $\lambda 5413$) is an effective optical depth defined as the following:
\begin{equation}
   \tau(\ion{He}{II}\,\,\lambda5413)(r)= \left(\dfrac{1}{\tau_\mathrm{Static}(r)} + \dfrac{1}{\tau_\mathrm{Sob}(r)} \right)^{-1}.
   \label{eq:tau_HeII_5413}
\end{equation}

\noindent
In this expression $\tau_\mathrm{Static}$ is the optical depth for \ion{He}{II} $\lambda 5413$ in a static atmosphere, $\tau_\mathrm{Sob}$ is the Sobolev optical depth for \ion{He}{II} $\lambda 5413$, and $r$ is the radius. For small $r$, the wind's velocity is small, leading $\tau$(\ion{He}{II} $\lambda 5413$) to approach $\tau_\mathrm{Static}$; at larger values of $r$, $\tau$(\ion{He}{II} $\lambda 5413$) will approach $\tau_\mathrm{Sob}$. The expression for $\tau$ is only used for diagnostic purposes.

\begin{figure*}  
	\begin{minipage}[t]{\linewidth}
		\centering
		\begin{subfigure}{.3\textwidth}
			%	\centering
			\includegraphics[width=1.07\linewidth]{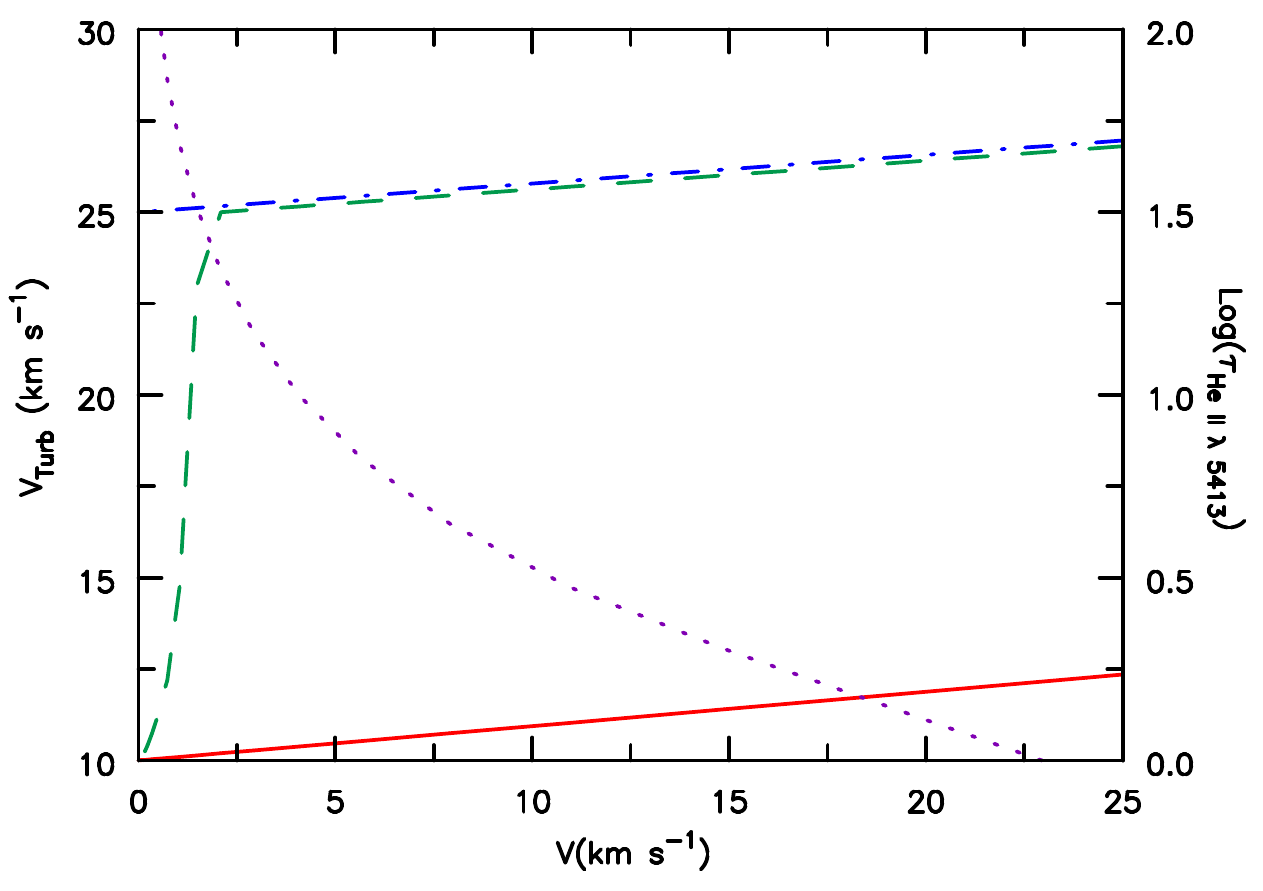}
					\end{subfigure}
		\begin{subfigure}{.30\textwidth}
			%	\centering
			\includegraphics[width=1\linewidth]{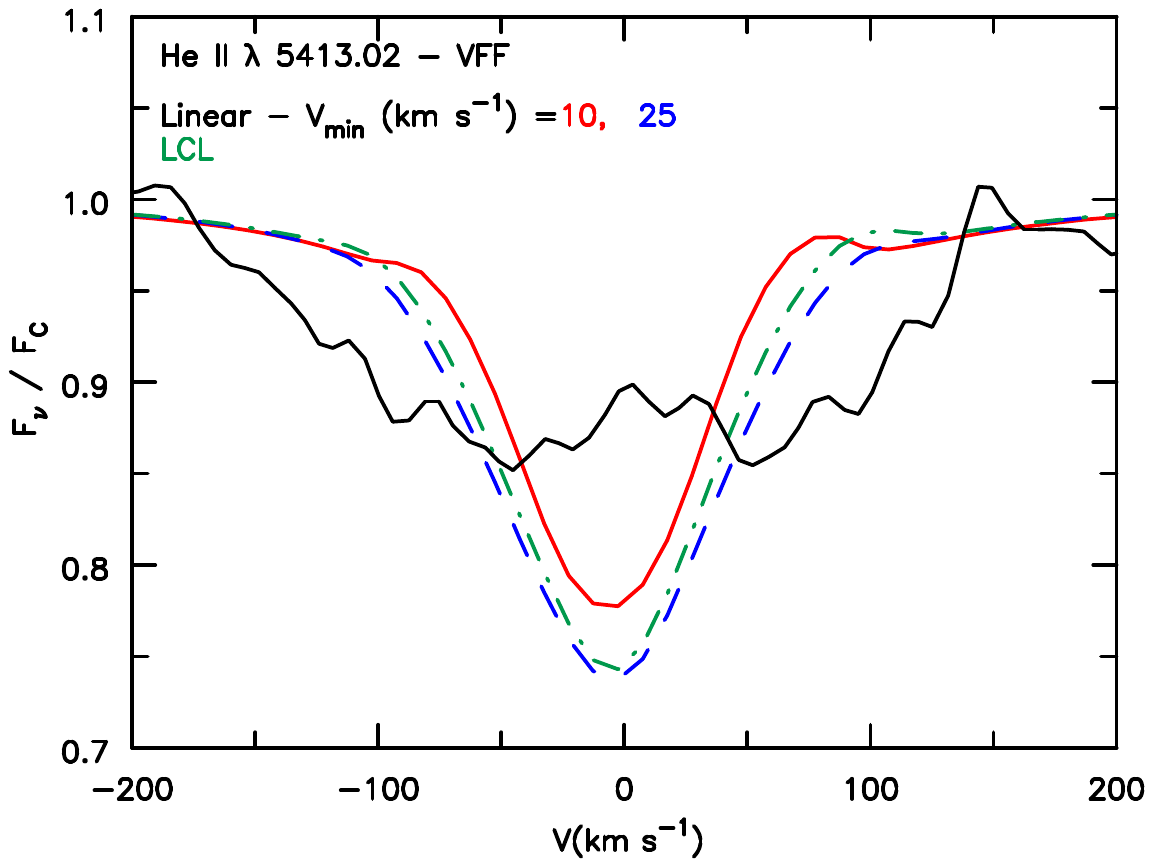}
		\end{subfigure}
		\begin{subfigure}{.3\textwidth}
			%\centering
			\includegraphics[width=1\linewidth]{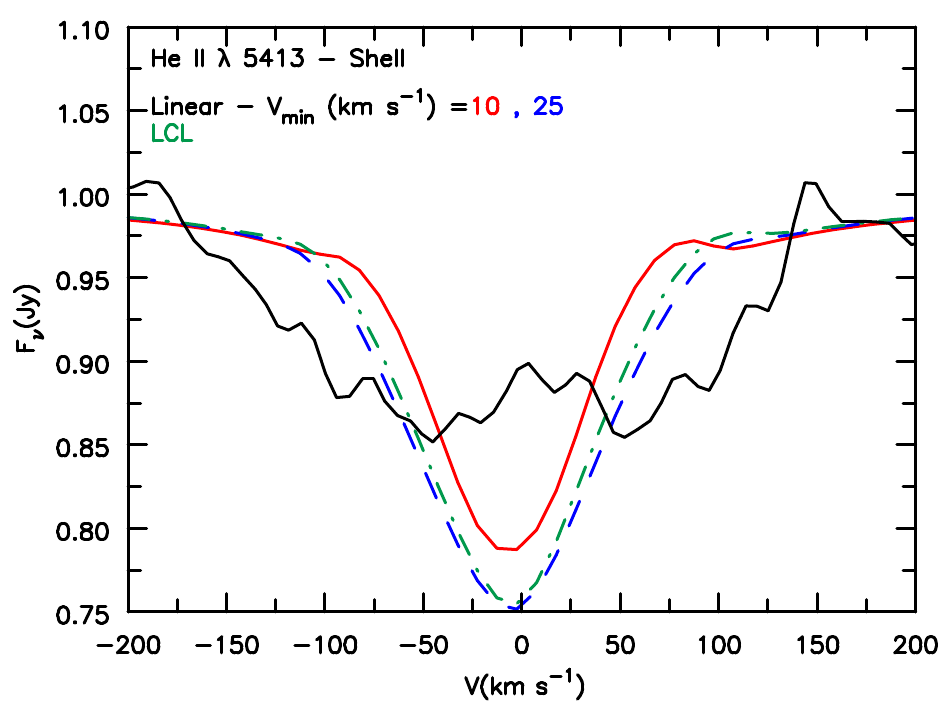}
		\end{subfigure}%
	\end{minipage}
	\caption{Illustration of the \textit{Linear-CubeRoot-Linear} (LCL) function as a possible description of depth-dependent turbulence, compared to the linear function from Eq. \ref{eq:vturb_law} \textit{(left)}. Two curves using the linear description of turbulence, $\varv_\mathrm{min}$= 10 \kms \textit{(red, solid curve)} and $\varv_\mathrm{min}$= 25 \kms \textit{(blue, dot-dash curve)}, are compared with the LCL description \textit{(green, dashed curve)}.   Placing the acceleration portion of the LCL function near $\log \tau$ (\ion{He}{II} $\lambda 5413$) = 1.0, \textit{(purple, dotted curve)}, we again synthesize \ion{He}{II} $\lambda 5413$ with the VFF \textit{(middle)} and Shell \textit{(right)} approach. \ion{He}{II} $\lambda 5413$ has increased in strength for both approaches, but neither replicates its  broad profile.}
	\label{fig_LCL}
\end{figure*}

In Fig. \ref{fig_LCL}, we  plot \ion{He}{II} $\lambda$5413 using the LCL description of turbulence, and 
compare it with both observation, and profiles computed using  the usual turbulence description.
As apparent, we are unable to reproduce the breadth of the line. Further studies of these lines should be conducted since they may reveal valuable information about the structure of the wind around the sonic point.

\section{Discussion}\label{sec_discuss}

When implementing any treatment for clumping in a radiative transfer code, we should be mindful of the  physical assumptions made and their influence on our results. The VFF method has been successful in treating winds with clumping but its assumptions are too restrictive. The Shell method has several important strengths that are irreproducible with the VFF method and it is worthwhile to discuss these. 

As described in Section \ref{sec_intro}, when implementing the VFF method, we make the assumption that the clumpy wind is described by a homogeneous distribution of small-scale clumps embedded in a void background. The clumps have a density $\rho/f_\mathrm{v}$ where $\rho$ is the local smooth wind's density defined by 
$$\rho=\Mdot /4\pi r^2 v \,\,. $$ 
Thus the relations between the density of the clumped wind ($\rho_\mathrm{clw}(r)$),  $\rho(r)$, and $f_\mathrm{v}$ are
\begin{equation*}
\hspace{1cm}  \langle \rho_\mathrm{clw} \rangle = \langle\rho\rangle \quad \textrm{and} \quad \langle {\rho_\mathrm{clw}}^2 \rangle = \langle \rho^2 \rangle/f_\mathrm{v}\,\,,
  \label{eq:rho}
\end{equation*}
where $<>$ denotes a spatial average. As apparent from these scalings, density-dependent features will remain similar, while density squared features are enhanced by a factor of $1/f_\mathrm{v}$ for a given mass-loss rate.

However, our treatment in \textsc{cmfgen} is necessarily only valid if all clumps are optically thin  -- that is, the mean free path of the photon is large compared to the size of the clumps, and there are ``many" clumps per mean free path. When these assumptions are invalid the transfer of photons will be affected by the size and geometry of the clumps, and their distribution. Due to the influence of the wind's velocity field on line transfer, which reduces the
size of the line interaction region to $\sim rV_\mathrm{Dop}/V(r)$, the VFF approach is likely to be more accurate for the continuum than for lines.

With the Shell method, we make no such assumptions and allow the opacity, and subsequently the optical depth, to be calculated from the  supplied density and velocity profile. Additionally, it incorporates a non-void ICM in the wind. This assumption, along with the inclusion of X-rays, has a profound influence on the wind's ionisation structure and UV resonance lines. As shown in Figs. \ref{fig_IF} and \ref{fig_superion_pop}, the ICM's ionisation structure differs immensely from its neighbouring shells, and the number densities of O$\,^{5+}$ and N$^{4+}$ in the shells are similar to those in the ICM. This was first pointed out in a theoretical analysis by \cite{Zsargo2008}, and was verified when they imported the radiation field of a VFF-wind model into a diffuse smooth-wind model, which acted as their ICM, in order to fit \ion{O}{VI} $\lambda 1035$ of $\zeta$ Pup. In conjunction with our results, this shows that the ICM cannot be ignored when studying the  formation of \ion{O}{VI} and \ion{N}{V} resonance lines.

To incorporate the Shell method into a radiative transfer code requires additional free parameters --  FWHM of the shell profile, minimum velocity at which shells are onset, etc. -- which at face-value brings more complexity to treating clumpy winds as opposed to the simple, two free parameters needed in the VFF approach, $f_\mathrm{v}$ and $V_\mathrm{cl}$. However, with complexity comes more freedom to experiment with modeling clumpy winds; having only two parameters to globally describe clumping is often too restrictive, hindering us from exploring various wind models outside those described with the VFF approach. 

In addition to the wind models presented in this paper, we also explored winds with a larger number of shells, winds with differing density contrasts, and winds with an earlier shell onset.
Doubling the number of shells resulted in a slight enhancement (at most, a 2\% increase in the equivalent width [EW]) of the red-emission portion of UV resonance lines and H$\alpha$ emission. Decreasing the ICM density by a factor of
2 in the outer wind  (but by a smaller amount in the inner wind where the shells are less distinct)  enhances the H$\alpha$ emission EW by about 5\%. This result  is not unexpected since H$\alpha$ emission mostly originates in the inner wind, and lowering the ICM density there enhances the density in the adjacent shells in order to retain the same mass-loss rate. Changes occurring to UV resonance lines will depend on the line; for example, \nvdoub\  and \ion{C}{IV} $\lambda\lambda1548, 1550$ will remain the same while \sivdoub\ will see some strengthening in its P Cygni profile (i.e. stronger absorption and emission).  A similar effect occurs when lowering the onset velocity of  the shells, though to a much lesser extent.

Nevertheless, using shells as a description of clumpy winds of hot, massive stars has its own limitations. The most obvious limitation is highlighted by hydrodynamical simulations of radiatively driven winds, such as those seen in \cite{Owocki2008} and \cite{Sundqvist2018}. Initially, these simulations show the development of shell-like structures, but these quickly break up due to hydrodynamical instabilities. Additionally, these hydrodynamical instabilities may generate shocks that propagate through the stellar wind, heating the gas such that it emits soft X-rays. This would be particularly important at the interface of the ICM and dense shells, where stronger hydrodynamical instabilities may occur. We do not calculate this from first principles, rather we rely on an empirical fit where X-ray luminosity is roughly correlated with the stellar luminosity.

The Shell model, like the VFF model, does not consider the effects of porosity discussed in \cite{Oskinova2007} and \cite{Owocki2006}. To implement porosity in a large non-LTE model would require a 3D radiative transfer model which is computationally prohibitive. Monte-Carlo codes have been used to explore the influence of porosity for more restricted problems such as the formation of a UV P~Cygni profile \citep{Surlan2012,Sundqvist2010,Sundqvist2011}. Other researchers have opted to use simple wind models to simulate porosity, as mentioned in Section~\ref{sec_intro}.

While there has been considerable effort to incorporate the effects of porosity, there is still debate in the hot-star community on  how relevant porosity is for the formation of wind lines. \cite{Oskinova2007} showed that porous winds can reduce the strength of \ion{P}{V}\ resonance lines, requiring a smaller lowering of the mass-loss rate derived from $\rho^2$ diagnostic lines (i.e. H$\alpha$).
Meanwhile, \cite{Owocki2008} achieved a similar reduction in line strength using a modeled wind with a strongly non-monotonic velocity field. This reduction occurs from the introduction of gaps in velocity space -- termed velocity porosity, or \textit{vorosity} -- allowing light to pass by without interacting with the wind material. However, this effect only leads to a 10-20\% reduction in line strength. Nevertheless, these results warrant further investigation into their relevance for line formation to enable us to better characterize the structure of stellar winds.

In our approach, the Shell models do allow for velocity porosity -- that is, the shells only absorb at discrete velocities. The importance of this effect depends on the adopted microturbulent velocity, and on the transition under investigation. Transitions such as \phosdoub\ are sensitive since the transitions are not black, and because the density of P$^{4+}$ ions in the ICM is too low to produce significant absorption. On the other hand, the profiles of \nvdoub\ and \osixdoub\ will be less sensitive due to the importance of the ICM in producing their observed line profiles.

Additionally, the Shell method retains the characteristic line formation behaviour seen in the VFF method. For instance, \sivdoub\ and \phosdoub\ are important contrasting probes for effective temperature and clumping. In AzV83, P$^{4+}$ is the dominant ion in the entire wind, whereas S$^{3+}$ is only dominant in the outer wind (S$^{4+}$ is dominant in the inner wind).
 
\section{Conclusions}\label{sec_conclusion}

We have described two procedures used for modeling clumpy winds -- the commonly adopted VFF approach and our Shell approach. Using the 1D radiative transfer code \textsc{cmfgen} we have compared models, based on each of the two procedures, with observations of AzV83. By design both methods used the same atmospheric parameters but the mass-loss rate of the Shell model
was increased by $\sim$40\% in order to match the \ion{He}{II} $\lambda$4686 and H$\alpha$ emission lines.

Our work illustrates the significance of allowing for the ICM which is overlooked in the VFF approach. There is a stark contrast between the predicted ionisation fractions in the Shell approach and the VFF approach, and in the predicted resonance profiles of the UV super-ions. In the Shell model, the ionisation fraction in the ICM differs from that in the neighbouring clumps by as much as three orders of magnitude. However, despite the large density contrast between a shell and the ICM, the number densities of \ion{N}{V} and \ion{O}{VI} vary smoothly along the wind. This results in a stronger P~Cygni resonance profile than would be produced by the VFF approach for a given X-ray luminosity.  The reason for the smooth behaviour has been explained by \cite{Zsargo2008}.

The non-surprising dependence of the ionisation structure on density also highlights another issue with the VFF and (to a lesser extent) the Shell approach -- the clumps at a given radius  have the same density whereas we might expect a range of clump sizes and shapes.

Although the present study provides insights into the effects on the ionisation structure and spectral characteristics of AzV83 when treating clumpy winds with shells, it still suffers from a number of important limitations that must be addressed by future work. For example, how much of an effect does porosity have? In what regimes does the use of coherent shells introduce artefacts into the modeling that affect derived parameters? In what regimes does treating winds with the Shell approach perform worse than the VFF approach, and why? Finally, the clump approach is parameterised, and does not use a clump and velocity  description derived from radiation-hydrodynamical simulations. 

A paper applying the Shell method to a WN star is in preparation. In that paper we show that the lateral coherence of the shells can have a significant influence on the strength of optically thick He recombination lines in a WN star. In our simulations the shell thickness can be similar in size to the Sobolev length. In such cases the Sobolev approximation is invalid, and the blob structure will influence photon escape. The angle-dependent escape probability in asymmetric blobs, which will in general  be different from the angle-dependent Sobolev escape probability,  could, for example, explain the very rounded \ion{He}{ii}\  profiles seen in some broad-lined WN stars
\citep{2014A&A...562A.118S}.

In our Shell model we derived a mass-loss rate nearly 40\% higher than the VFF model, highlighting the extreme difficulty of deriving accurate mass-loss rates from spectral fitting. Accurate derivation of mass-loss rates will require more realistic hydrodynamical simulations, more realistic radiative transfer simulations, and the incorporation of rotation.

In general, both approaches produce synthetic spectra for AzV83 that are in reasonable agreement with observation. However, some significant discrepancies remain which need to be addressed by future studies. For example, why do our models consistently predict  excess red emission in UV resonance lines, particularly \sivdoub? Additionally, why can none of the models explain the observed absorption profile of \ion{He}{II} $\lambda$5413?

\section*{Acknowledgements}

BLF and DJH acknowledge support from STScI theory grant  HST-AR-14568.001-A. STScI is operated by the Association of Universities for Research in Astronomy, Inc., under NASA contract NAS 5-26555.  We thank Philip Massey for supplying an additional optical spectrum of AzV83, and the referee for providing useful comments which improved the manuscript.

The authors thank the referee for their careful reading of the manuscript, and their thoughtful comments.

\section*{Data Availability}
The UV data used in this study is available via the HST data archive. Optical data will be made available on request. An earlier version of \cmfgen, and the associated atomic data, is available at www.pitt.edu/\verb+~+hillier. Updates of this website are routinely made. Models underlying this article will be shared on reasonable request to the corresponding author.

%%%%%%%%%%%%%%%%%%%%%%%%%%%%%%%%%%%%%%%%%%%%%%%%%%


%\nocite{*}
\bibliographystyle{mnras}
\bibliography{shells_bib,atomic_bib}
%%%%%%%%%%%%%%%%%%%%%%%%%%%%%%%%%%%%%%%%%%%%%%%%%%

%%%%%%%%%%%%%%%%% APPENDICES %%%%%%%%%%%%%%%%%%%%%

%\FloatBarrier 

\appendix

\section{Atomic Models and  Data}
\label{atomic_data}

The model atoms used in the calculations are described in Table \ref{table:atom}. In Table \ref{table:atom}, $N_\mathrm{F}$ refers to the total number of levels included in the atom and $N_\mathrm{S}$ refers to the number of superlevels used in each atom. Additional atoms that were used but not included in the table include: \ion{Ne}{II-IV}, \ion{Cl}{IV-VI}, \ion{Ar}{III-V}, \ion{Ca}{III-VI}, \ion{Cr}{IV-VI}, \ion{Mn}{IV-VI}, and \ion{Ni}{IV-VI}. These atoms do not strongly affect the predicted spectrum.

\begin{table}%[h]
	\centering
	\caption{Summary of model atoms for several species.}
	\label{table:atom}
	\begin{tabularx}{\columnwidth}{*{5}{X}} 
		\hline
		&\textbf{Species} & \textbf{$N_\mathrm{S}$}  &\textbf{$N_\mathrm{F}$} &
		\\ \hline 
		
		&\ion{H}{I}       &  20  &  30 &    \\
		&\ion{He}{I}      &  45  &  45 &    \\
		&\ion{He}{II}     &  22  &  30 &    \\
		&\ion{C}{II}      &  14  &  14 &    \\
		&\ion{C}{III}     &  30  &  54 &    \\
		&\ion{C}{IV}      &  14  &  18 &  \\
		&\ion{N}{III}     &  34  &  70 &    \\
		&\ion{N}{IV}      &  29  &  53 &    \\
		&\ion{N}{V}       &  14  &  21 &    \\
		&\ion{O}{III}     &  25  &  25 &    \\
		&\ion{O}{IV}      &  23  &  49 &    \\
		&\ion{O}{V}       &  41  &  78 &    \\
		&\ion{O}{VI}      &  44  &  50 &    \\
		&\ion{Si}{III}    &  20  &  34 &    \\ 
		&\ion{Si}{IV}     &  23  &  33 &    \\
		&\ion{P}{IV}      &  36  & 178 &    \\
		&\ion{P}{V}       &  16  &  62 &    \\
		&\ion{S}{III}     &  15  &  28 &    \\
		&\ion{S}{IV}      &  57  & 142 &    \\
		&\ion{S}{V}       &  60  &  98 &    \\
		&\ion{S}{VI}      &  37  &  58 &    \\
		&\ion{Fe}{IV}     & 100  &1000 &    \\
		&\ion{Fe}{V}      &  61  & 300 &    \\
		&\ion{Fe}{VI}     & 133  & 444 &    \\
		&\ion{Fe}{VII}    &  78  & 153 &    \\ \hline
	\end{tabularx}
\end{table}

Atomic data used in the calculations is routinely updated, and can be downloaded from \url{http:/www.pitt.edu/~hillier}. Multiple data sets are available for many species. Sources for the atomic data used in the current paper are as follows:

  C\,{\sc iv} oscillator strengths for $n < 6$ are from \cite{LB11_Li_like}.
  Other values are from \cite{Lei72_CIV} ($n < 8$) or are hydrogenic.
  C\,{\sc iv} photoionisation data is from \cite{PSS88_LI_seq} and \cite{Lei72_CIV}, with
  hydrogenic rates used for high $l$ and $n$ states.
  Collision strengths for C\,{\sc iv} for $n < 6$ are from \cite{LB11_Li_like}.
  C\,{\sc iii} oscillator strengths, photoionisation data, and dielectronic line data  are from P.~J. Storey (private communication). Collision rates among the lowest six terms are from \cite{BBD85_col}.
  C\,{\sc ii} oscillator strengths  and photoionisation cross-sections are from \cite{DSK00_CII_recom}
              (\& P.~J. Storey private communication).
  NIST \citep{NIST_V5}  oscillator values are used when available.  
  Collision rates are from \cite{Tay08_CII_col}.
  
  N\,{\sc v} oscillator strengths for $n < 6$ are from \cite{LB11_Li_like},
  photoionisation data is from \cite{PSS88_LI_seq} with 
  hydrogenic rates used for high $l$ and $n$ states, and 
  collision strengths are from \cite{LB11_Li_like}.
  
  N\,{\sc iv} oscillator strengths and photoionisation cross-sections were computed by \cite{TSB90_Be_seq} and were obtained from TOPbase \citep{Topbase93}. The
  identification of the 2s\,4s\,$^3$S and 2p\,3p\,$^3$S states have been switched \citep[see][]{AAL91_NIV}.
  Oscillator strengths for N\,{\sc iv} forbidden (and semi-forbidden) lines are from \cite{NS79_Inter}.
  N\,{\sc iv}  oscillator strengths for doubly excited states above the
  ionisation limit are  from \cite{NS83_LTDR,NS84_CNO_LTDR}.
   
  N\,{\sc iii} oscillator strengths photoionisation cross-sections were obtained from TOPbase \citep{Topbase93}.
  Transition probabilities for the N\,{\sc iii} intercombination line are from \cite{NS79_NIII} and these
  are in reasonable agreement with experimental values found study of \cite{TGK99_CII_NIII}.
  Collision strengths for transitions amongst the first 20 levels are from \cite{SBH94_NIII_col}.
  
  O\,{\sc vi} oscillator strengths for $n < 6$ are from \cite{LB11_Li_like}, photoionisation data is
  from \cite{PSS88_LI_seq}, and collision strengths are from \cite{LB11_Li_like}.
   O\,{\sc v} oscillator strengths are from \cite{NS83_LTDR,NS84_CNO_LTDR} (private communication).
    Collision rates for the six lowest terms of O\,{v} are from \cite{BBD85_col}.
 O\,{\sc iv} oscillator strengths were  obtained from TOPbase \citep{Topbase93}.
  Intercombination data ($^2$P-$^4$P) for O\,{\sc iv} is from
  the compilation of \cite{Men83_col}.
  Photoionisation data is from \cite{Nah98_Ophot} and
  were obtained through NORAD \citep{Nahar_NORAD}.
  O\,{\sc iv} collision rates are from \cite{ZCO94_B_like}.
 O\,{\sc iii} oscillator strengths and photoionisation data are from \cite{LPS89_OIII_phot}.
  Collision strengths  are from Keith Butler (private communication, 2012)
  
  Oscillator strengths for Ne\,{\sc iv} are from the compilation by \cite{KP75_linelist}
  and the opacity project \citep{Sea87_OP}. Photoionisation cross-sections were obtained from TOPbase \citep{Topbase93}.
  Collisional data for the three lowest terms of Ne\,{\sc iv} is from the compilation of \cite{Men83_col}.
Oscillator strengths for Ne\,{\sc iii } are from the compilation by \cite{KP75_linelist}
  and the opacity project \citep{Sea87_OP}. Photoionisation cross-sections were obtained from TOPbase \citep{Topbase93}.

  Si\,{\sc iv} oscillator strengths and photoionisation cross-sections  were obtained from TOPbase \citep{Topbase93}. 
  Oscillator strengths have been updated with  values from \cite{NIST_V5} (2012-10-14).
  Si\,{\sc iv} collision strengths (for $n < 7$) are from \cite{LWB09_Na_seq_col}.
 Si\,{\sc iii} photoionisation cross-sections were computed by \cite{BKM93_Mg_seq} and were obtained from TOPbase \citep{Topbase93}. Transition probabilities have been updated with  values from \cite{NIST_V5}. 
  
   Photoioionisation cross-sections for S\,{\sc vi} are from  \cite{1987JPhB...20.3899D}.
   S\,{\sc iv} oscillator strengths and photoionisation cross-sections were retrieved from TOPbase  \citep{Topbase93}. Intercombination data for S\,{\sc iv} is from the compilation of \cite{Men83_col}. Transition probabilities have been updated using the critical compilation by \cite{NIST_V5p2}.
 S\,{\sc iii} oscillator strengths and photoionisation cross-sections  were computed by \cite{NP93_Sil} and were obtained from \cite{Topbase93}. Data for forbidden and intercombination transitions are from \cite{Tay97_SIII_forb}, \cite{Men83_col}, and \cite{Hua85_Si_seq}. Transition probabilities have been updated using the critical compilation by \cite{NIST_V5p2}.  Collision strengths for S\,{iii} are from \cite{TG99_SIII_col}.
 
 \begin{table}
 	\centering
 	\caption{Summary of abundances of several species.}
 	\label{table:abund}
 	\begin{tabularx}{\columnwidth}{ *{4}{X} }
 		\hline
 		&\textbf{Abundance} &  &		\\ 
 		\hline \\[-4pt]
 		
 		&$\frac{N(\text{He})}{N(\text{H})}$       & 0.2   &   \\[5pt]
 		&$X(\text{He})$                           & 0.44  &   \\[5pt]
 		
 		&$\frac{N(\text{C})}{N(\text{H})}$          & $3.76\times 10^{-5}$ &\\[5pt]
 		&$\frac{X(\text{C})}{X(\text{C})_\mathrm{\odot}}$  & $8.19\times 10^{-2}$ &\\[5pt]
 		
 		&$\frac{N(\text{N})}{N(\text{H})}$          & $2.58\times 10^{-4}$ &\\[5pt]
 		&$\frac{X(\text{N})}{X(\text{N})_\mathrm{\odot}}$  & $1.83$               &\\[5pt]
 		
 		&$\frac{N(\text{O})}{N(\text{H})}$          & $6.32\times 10^{-5}$ &\\[5pt]
 		&$\frac{X(\text{O})}{X(\text{O})_\mathrm{\odot}}$  & $5.87\times 10^{-2}$ &\\[5pt]
 		
 		&$\frac{X(\text{Si})}{X(\text{Si})_\mathrm{\odot}}$ & $0.203$ &\\[5pt]
 		
 		&$\frac{X(\text{P})}{X(\text{P})_\mathrm{\odot}}$  & $5.80\times 10^{-2}$ &\\[5pt]
 		
 		&$\frac{X(\text{S})}{X(\text{S})_\mathrm{\odot}}$  & $0.268$ &\\[5pt]
 		
 		&$\frac{X(\text{Fe})}{X(\text{Fe})_\mathrm{\odot}}$  & $0.185$& \\[5pt]
 		
 		\hline\\
 	\end{tabularx}
 \end{table}

  Oscillator strengths for Fe\,{iv}, Fe\,{v}, Fe\,{vi}, and Fe\,{\sc vii} were computed by Bob Kurucz \citep{Kur09_ATD} and were obtained (prior to 2001) from his website \citep{Kur_web}.  Photoionisation cross-sections for these species were computed as part of the opacity project   \citep{Sea87_OP} and were obtained from TOPbase \citep{Topbase93}. Collision strengths for Fe\,{\sc vi} are from \cite{CP99_FeVI_col}.  Fe\,{\sc iv} collisional data is from \cite{ZP97_FeIV_col}.

%\FloatBarrier
\section{Summary of Abundances}
\label{abundances}

 Table \ref{table:abund} lists the abundances of the species used in the model calculations where N(A) and X(A) denote the relative number fraction and mass fraction of species A, respectively.

\section{Non-LTE Effects}
\label{Non-LTE Effects}

\begin{figure}
	\includegraphics[width=8.5cm]{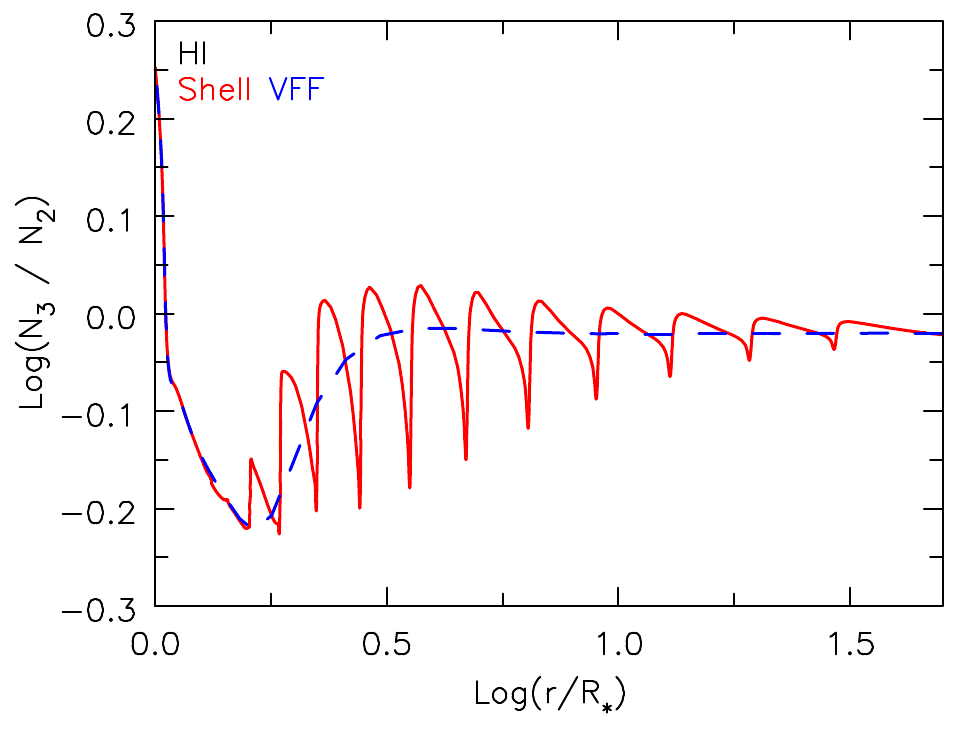}
	\caption{We plot the ratio level population of hydrogen, specifically $N_{3}$ on $N_{2}$, in a Shell \textit{(solid curve)} and a VFF \textit{(dashed curve)} model. The number density of the VFF model varies smoothly with increasing radius while the Shell model's does not. While the Shell model on average follows the general trend seen in the VFF model, $N_{3}$ on $N_{2}$ is overall greater in the ICM while within the shell it is lower compared to the VFF model. 
	This characteristic arises from the enhanced recombination rate seen within the shell. 
	The populations in the Shell model approach the values  in the VFF model in the outer wind. }
	\label{fig_num_den_ratio}
\end{figure}

Using shells to represent clumps in stellar winds presents an excellent opportunity to examine some of the possible non-LTE effects that are not observed in the VFF approach. In Fig. \ref{fig_num_den_ratio} we show  the  $n_{3}$ to $n_{2}$  population ratio for the Shell and VFF models. In the VFF model, this ratio varies smoothly with increasing radius, however in the Shell model the variation is much more complicated. While it follows the general trend seen in the VFF model, $n_{3}$ on $n_{2}$ is lower in the shell compared to the ICM. This behaviour arises because
of enhanced recombination which boosts the level two population.

\begin{figure} 
	\includegraphics[width=8.2cm]{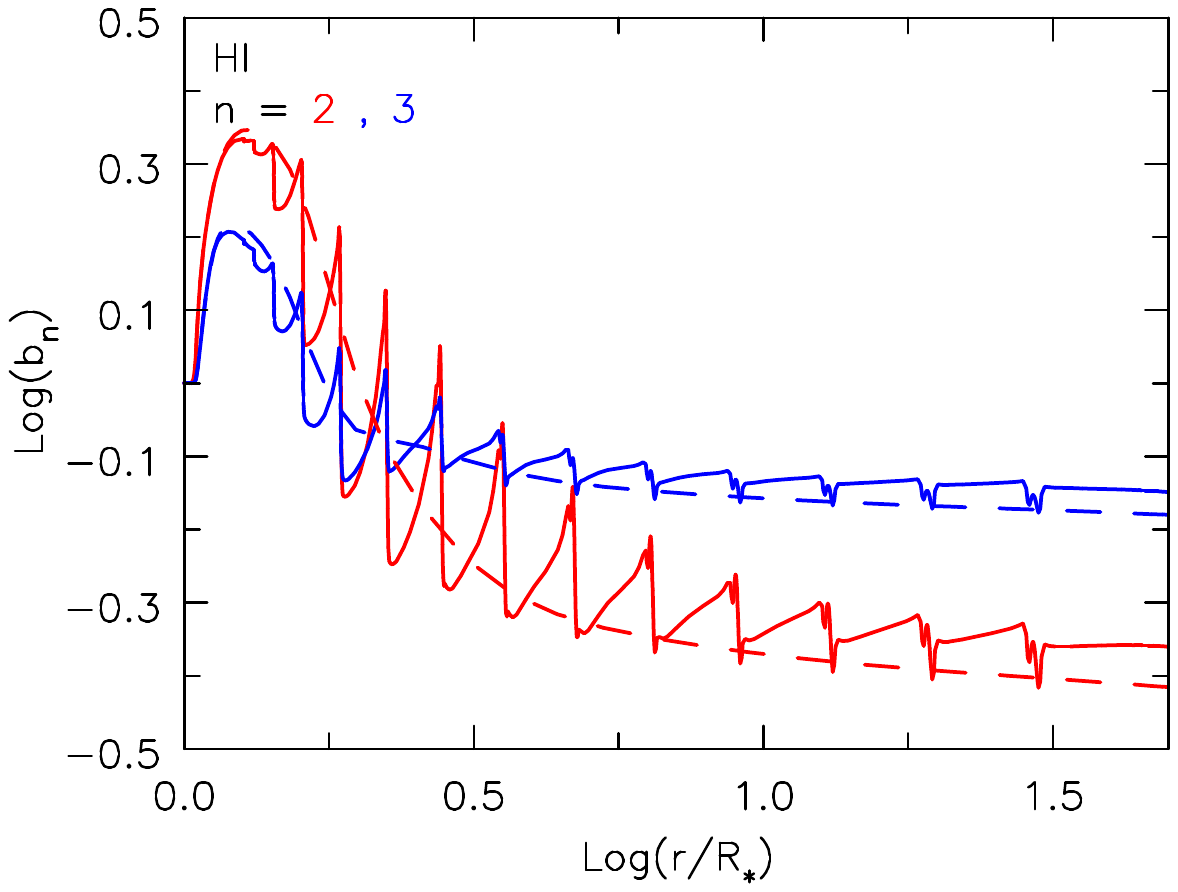}
	\includegraphics[width=8.5cm]{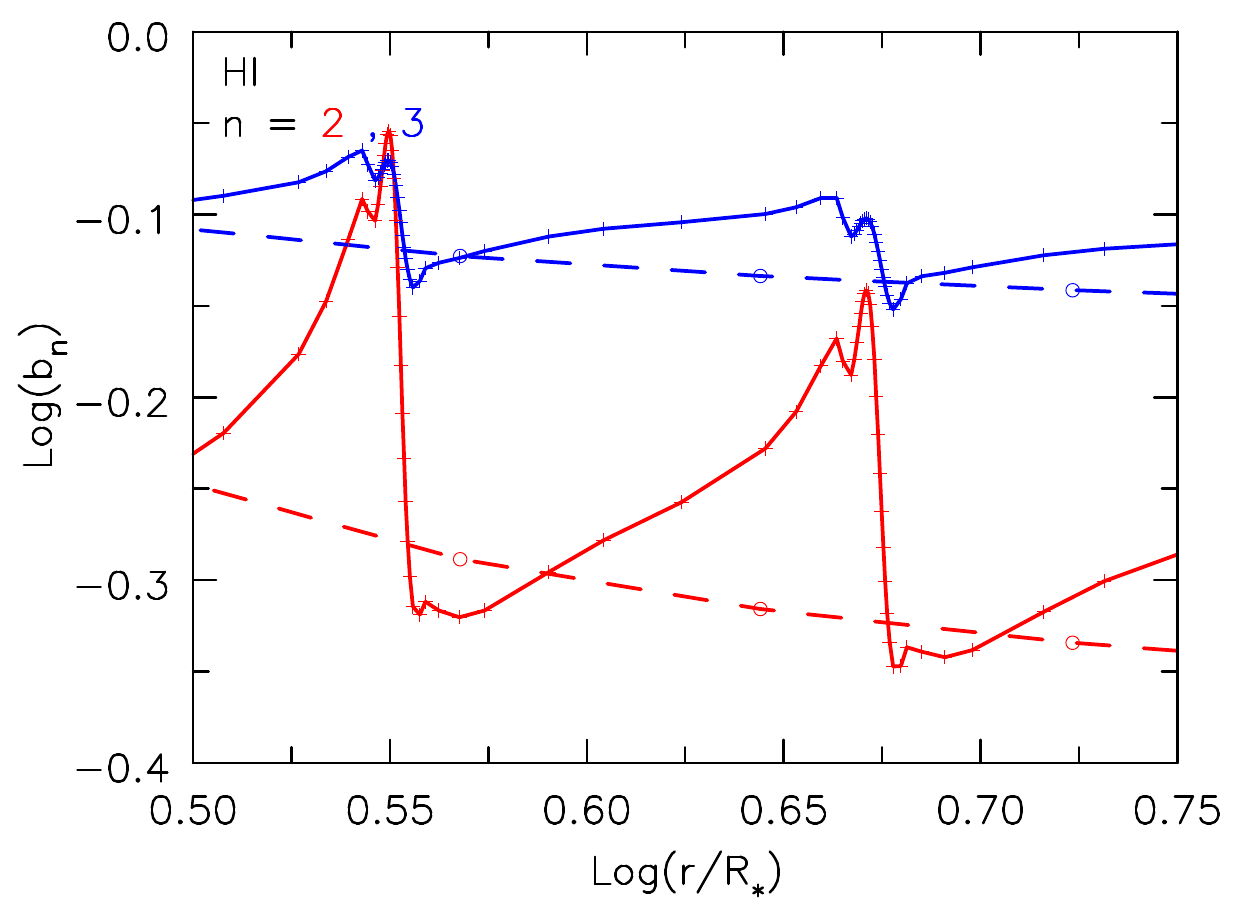}
	\caption{We plot (\textit{top}) the departure coefficient (DC) of hydrogen, specifically $n_{2}$ and $n_{3}$, in a Shell \textit{(solid curve)} and a VFF \textit{(dashed curve)} model. The bottom plot shows a closeup of the DC of $n_{2}$ and $n_{3}$ across two shells. Here we see again the smooth change in DC of $n_{2}$ and $n_{3}$ in the VFF model, whereas the Shell model shows a more complex variation throughout the wind. More interestingly, the shells themselves show further variations in the DCs of hydrogen. }
	\label{fig_hi_dc}
\end{figure}

In Fig. \ref{fig_hi_dc} we present the departure coefficient (DC) of hydrogen for $n_{2}$ and $n_{3}$ in a Shell and in a VFF model. Similar to what was already seen in Fig. \ref{fig_num_den_ratio} we see that in a VFF model the DC varies smoothly with radius, while the variation in the Shell mode is much more complicated. The changes are much
larger for the $n=2$ population and the DC varies across the shell. This is not unexpected, since, for example,
the finite width of a shell means that the escape probability of a line will depend on the location in the shell, 
and the radiation field is not isotropic.

The variations seen here in the Shell model highlight the complicated bahaviour that can occur in
a real clumped medium in which there is a complicated distribution of clump size, shapes, and
velocity widths. Clearly, studies with more realistic clumping distributions are warranted.

%%%%%%%%%%%%%%%%%%%%%%%%%%%%%%%%%%%%%%%%%%%%%%%%%%

% Don't change these lines
\bsp	% typesetting comment
\label{lastpage}

%%%%%%%%%%%%%%%%%%%%%%%%%%%%%%%%%%%%%%%%%%%%%%%%%%%%%%%%%%%%%%%%%%%%

\end{document}